\documentclass[a4paper]{report}
\usepackage[utf8]{inputenc}
\usepackage[T1]{fontenc}
\usepackage{RJournal}
\usepackage{amsmath,amssymb,array}
\usepackage{booktabs}
\usepackage{enumitem}

\usepackage{bm}
\usepackage[ruled, vlined]{algorithm2e}

\begin{document}

\sectionhead{}
\volume{}
\volnumber{}
\year{}
\month{}

\begin{article}

\title{\pkg{fnets}: An R Package for Network Estimation and Forecasting via Factor-Adjusted VAR Modelling}
\author{by Dom Owens, Haeran Cho and Matteo Barigozzi}

\maketitle

\abstract{
The package \CRANpkg{fnets} for the R language implements the suite of methodologies proposed by \cite{barigozzi2022fnets} for the network estimation and forecasting of high-dimensional time series under a factor-adjusted vector autoregressive model, which permits strong spatial and temporal correlations in the data.
Additionally, we provide tools for visualising the networks underlying the time series data after adjusting for the presence of factors.
The package also offers data-driven methods for selecting tuning parameters including the number of factors, order of autoregression and thresholds for estimating the edge sets of the networks of interest in time series analysis.
We demonstrate various features of \pkg{fnets} on simulated datasets as well as real data on electricity prices.
}

\section{Introduction}

Vector autoregressive (VAR) models are popularly adopted for modelling time series datasets collected in many disciplines including economics \citep{koop2013forecasting}, finance \citep{barigozzi2019nets}, neuroscience \citep{kirch2015eeg} and systems biology \citep{shojaie2010discovering}, to name a few.
By fitting a VAR model to the data, we can infer dynamic interdependence between the variables and forecast future values.
In particular, estimating the non-zero elements of the VAR parameter matrices recovers directed edges between the components of vector time series in a Granger causality network.
Besides, by estimating the precision matrix (inverse of the covariance matrix) of the VAR innovations, we can define a network representing their contemporaneous dependencies by means of partial correlations.
Finally, the inverse of the long-run covariance matrix of the data simultaneously captures lead-lag and contemporaneous co-movements of the variables. 
For further discussions on the network interpretation of VAR modelling, we refer to
\cite{dahlhaus2000graphical}, \cite{eichler2007granger}, \cite{billio2012econometric} and \cite{barigozzi2019nets}.

\begin{figure}[htb!]
\centering
\includegraphics[width = .6\textwidth]{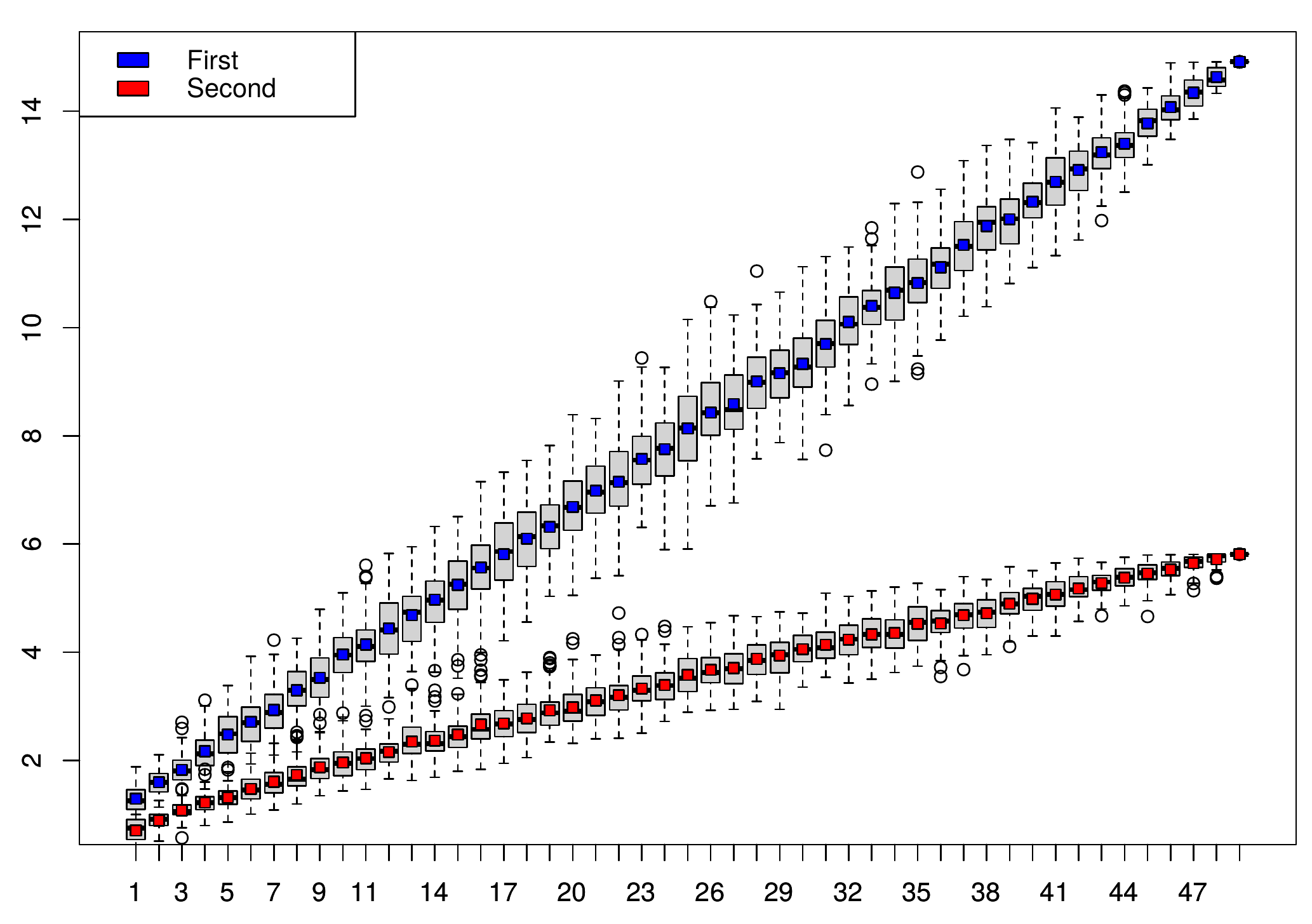}
\caption{Box plots of the two largest eigenvalues ($y$-axis) of the long-run covariance matrix estimated from the energy price data collected between 01/01/2021 and 19/07/2021 ($n = 200$), see \hyperref[sec:real]{Data example} for further details.
Cross-sections of the data are randomly sampled $100$ times for each given dimension $p \in \{2, \dots, 50\}$ ($x$-axis) to produce the box plots.}
\label{fig:eigen}
\end{figure}

\begin{figure}[htb!]
\centering
\includegraphics[width = .4\textwidth]{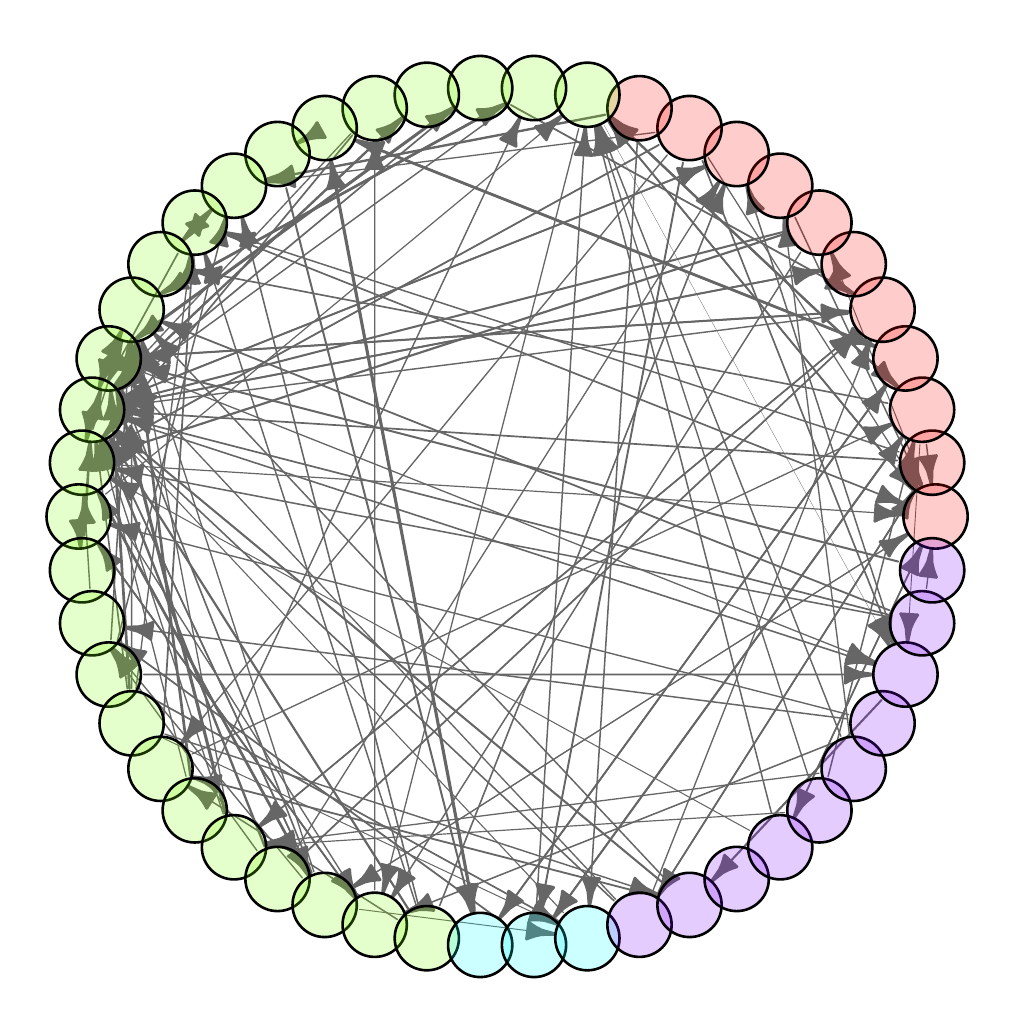}
\includegraphics[width = .4\textwidth]{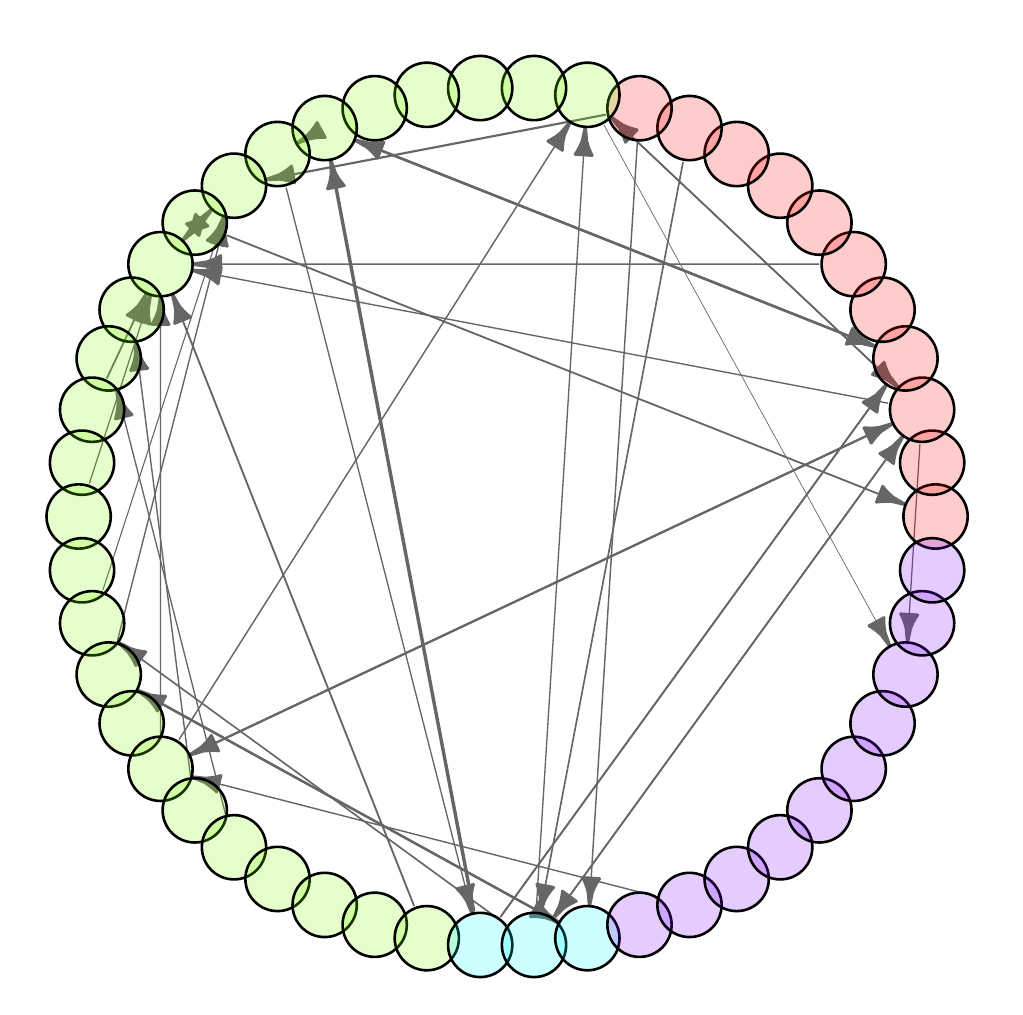}
\caption{Granger causal networks defined in~\eqref{eq:net:dir} obtained from fitting a VAR($1$) model to the energy price data analysed in Figure~\ref{fig:eigen}, without (left) and with (right) the factor adjustment step outlined in \hyperref[sec:estimation]{FNETS: Network estimation}. 
Edge weights (proportional to the size of coefficient estimates) are visualised by the width of each edge, and the nodes are coloured according to their groupings, see \hyperref[sec:real]{Data example} for further details.}
\label{fig:q0q1}
\end{figure}

Fitting VAR models to the data quickly becomes a high-dimensional problem
as the number of parameters grows quadratically with the dimensionality of the data.
There exists a mature literature on $\ell_1$-regularisation methods for estimating VAR models in high dimensions under suitable sparsity assumptions on the VAR parameters
\citep{basu2015regularized, han2015direct, kock2015oracle, medeiros2016, nicholson2020high, liu2021robust}.
Consistency of such methods is derived under the assumption that the spectral density matrix of the data has bounded eigenvalues.
However, in many applications, the datasets exhibit strong serial and cross-sectional correlations which leads to the violation of this assumption.
As a motivating example, we introduce a dataset of node-specific prices in the PJM (Pennsylvania, New Jersey and Maryland) power pool area in the United States, see \hyperref[sec:real:energy]{Energy price data} for further details.
Figure~\ref{fig:eigen} demonstrates that the leading eigenvalue of the long-run covariance matrix (i.e.\ spectral density matrix at frequency $0$) increases linearly as the dimension of the data increases, which implies the presence of latent common factors in the panel data \citep{forni2000generalized}.
Additionally, the left panel of Figure~\ref{fig:q0q1} shows the inadequacy of fitting a VAR model to such data under the sparsity assumption via $\ell_1$-regularisation methods, unless the presence of strong correlations is accounted for by a {\it factor-adjustment} step as in the right panel.

\cite{barigozzi2022fnets} propose the FNETS methodology for factor-adjusted VAR modelling of high-dimensional, second-order stationary time series.
Under their proposed model, the data is decomposed into two latent components such that the {\it factor-driven} component accounts for pervasive leading, lagging or contemporaneous co-movements of the variables, while the remaining {\it idiosyncratic} dynamic dependence between the variables is modelled by a sparse VAR process.
Then, FNETS provides tools for inferring the networks underlying the latent VAR process and forecasting.

In this paper, we present an R package named \pkg{fnets} which implements the FNETS methodology.
It provides a range of user-friendly tools for estimating and visualising the networks representing the interconnectedness of time series variables, and for producing forecasts.
In addition, \pkg{fnets} thoroughly addresses the problem of selecting tuning parameters ranging from the number of factors and the VAR order, to regularisation and thresholding parameters adopted for producing sparse and interpretable networks.
As such, a simple call of the main routine of \pkg{fnets} requires the input data only, and it outputs an object of S3 class \code{fnets} which is supported by a \code{plot} method for network visualisation and a \code{predict} method for time series forecasting.

There exist several packages for fitting VAR models and their extensions to high-dimensional time series, see \CRANpkg{lsvar} \citep{lsvar}, \CRANpkg{sparsevar} \citep{sparsevar}, \CRANpkg{nets} \citep{nets}, \CRANpkg{mgm} \citep{haslbeck2020mgm}, \CRANpkg{graphicalVAR} \citep{epskamp2018gaussian}, \CRANpkg{bigVAR} \citep{nicholson2017bigvar},
and \CRANpkg{bigtime} \citep{wilms2021bigtime}.
There also exist R packages for time series factor modelling such as \CRANpkg{dfms} \citep{dfms} and \CRANpkg{sparseDFM} \citep{mosley2023sparsedfm}, and \CRANpkg{FAVAR} \citep{bernanke2005measuring} for Bayesian inference of factor-augmented VAR models.
The package \CRANpkg{fnets} is clearly distinguished from, and complements, the above list by handling strong cross-sectional and serial correlations in the data via factor-adjustment step performed in frequency domain.
In addition, the FNETS methodology operates under the most general approach to high-dimensional time series factor modelling termed the Generalised Dynamic Factor (GDFM), first proposed in \cite{forni2000generalized} and further investigated in \cite{forni2015dynamic}.
Accordingly, \pkg{fnets} is the first R package to provide tools for high-dimensional panel data analysis under the GDFM, such as fast computation of spectral density and autocovariance matrices via the Fast Fourier Transform, but it is flexible enough to allow for more restrictive static factor models.
While there exist some packages for network-based time series modelling (e.g.\ \CRANpkg{GNAR}, \citeauthor{knight2020generalized}, \citeyear{knight2020generalized}), we highlight that the goal of \CRANpkg{fnets} is to learn the networks underlying a time series and does not require a network as an input.

\section{FNETS methodology}
\label{sec:models}

In this section, we introduce the factor-adjusted VAR model and describe the FNETS methodology proposed in \cite{barigozzi2022fnets} for network estimation and forecasting of high-dimensional time series.
We limit ourselves to describing the key steps of FNETS and refer to the above paper for its comprehensive treatment, both methodologically and theoretically. 

\subsection{Factor-adjusted VAR model}

A zero-mean, $p$-variate process $\bm\xi_t$ follows a VAR($d$) model if it satisfies
\begin{align}
\label{eq:idio:var}
    \bm\xi_t &= \sum_{\ell = 1}^{d} \mbf A_{\ell} \bm\xi_{t - \ell} + \bm\Gamma^{1/2} \bm\vep_t,
\end{align}
where $\mbf A_\ell \in \R^{p \times p}, \, 1 \le \ell \le d$, determine how future values of the series depend on their past. 
For the $p$-variate random vector $\bm\vep_t = (\vep_{1t}, \ldots, \vep_{pt})^\top$, we assume that $\vep_{it}$ are independently and identically distributed (i.i.d.) for all $i$ and $t$ with $\E(\vep_{it}) = 0$ and $\Var(\vep_{it}) = 1$. Then, the positive definite matrix $\bm\Gamma \in \R^{p \times p}$ is the covariance matrix of the innovations $\bm\Gamma^{1/2} \bm\vep_t$.
 
In the literature on factor modelling of high-dimensional time series, the factor-driven component exhibits strong cross-sectional and/or serial correlations by `loading' finite-dimensional vectors of factors linearly.
Among many time series factor models, the GDFM \citep{forni2000generalized} provides the most general approach where the $p$-variate factor-driven component $\bm\chi_t$ admits the following representation
\begin{align}
    \bm\chi_t &= \mc B(L) \mbf u_t = \sum_{\ell = 0}^\infty \mbf B_\ell \mbf u_{t - \ell} \text{ \ with \ } \mbf u_t = (u_{1t}, \ldots, u_{qt})^\top \text{ and } \mbf B_\ell \in \R^{p \times q}, \label{eq:gdfm}
\end{align}
for some fixed $q$, where $L$ stands for the lag operator. 
The $q$-variate random vector $\mbf u_t$ contains the common factors which are loaded across the variables and time by the filter $\mc B(L) = \sum_{\ell = 0}^\infty \mbf B_\ell L^\ell$, and it is assumed that $u_{jt}$ are i.i.d.\ with $\E(u_{jt}) = 0$ and $\Var(u_{jt}) = 1$.
The model~\eqref{eq:gdfm} reduces to a static factor model \citep{bai2003, stock2002forecasting, fan2013large}, when $\mc B(L) = \sum_{\ell = 0}^s \mbf B_\ell L^\ell$ for some finite integer $s \ge 0$. Then, we can write 
\begin{align}
\label{eq:static}
    \bm\chi_t = \bm\Lambda \mbf F_t \text{ \ where \ } \mbf F_t = (\mbf u_t^\top, \ldots, \mbf u_{t - s}^\top)^\top \text{ \ and \ } \bm\Lambda = [\mbf B_0, \ldots, \mbf B_s]
\end{align}
with $r = q (s + 1)$ as the dimension of static factors $\mbf F_t$. 
Throughout, we refer to the models~\eqref{eq:gdfm} and~\eqref{eq:static} as {\it unrestricted} and {\it restricted} to highlight that the latter imposes more restrictions on the model.

\cite{barigozzi2022fnets} propose a factor-adjusted VAR model under which we observe a zero-mean, second-order stationary process $\mbf X_t = (X_{1t}, \ldots, X_{pt})^\top$ for $t = 1, \ldots, n$, that permits a decomposition into the sum of the unobserved components $\bm\xi_t$ and $\bm\chi_t$, i.e.\
\begin{align}
    \label{eq:model}
    \mbf X_t = \bm\xi_t + \bm\chi_t.
\end{align}
We assume that $\E(\vep_{it} u_{jt'}) = 0$ for all $i$, $j$, $t$ and $t'$ as is commonly assumed in the literature, such that $\E(\xi_{it} \chi_{i't'}) = 0$ for all $1 \le i, i' \le p$ and $t, t' \in \Z$.

\subsection{Networks}
\label{sec:networks}

Under~\eqref{eq:model}, it is of interest to infer three types of networks representing the interconnectedness of $\mbf X_t$ after factor adjustment. 
Let $\mc V = \{1, \ldots, p\}$ denote the set of vertices representing the $p$ cross-sections.
Then, the VAR parameter matrices, $\mbf A_\ell = [A_{\ell, ii'}, \, 1 \le i, i' \le p]$, encode the directed network $\mc N^{\dir} = (\mc V, \mc E^{\dir})$ representing Granger causal linkages, where the set of edges are given by
\begin{align}
\label{eq:net:dir}
\mc E^{\dir} = \l\{(i, i') \in \mc V \times \mc V: \, A_{\ell, ii'} \ne 0 \text{ for some } 1 \le \ell \le d \r\}.
\end{align}
Here, the presence of an edge $(i, i') \in \mc E^{\dir}$ indicates that $\xi_{i', t - \ell}$ Granger causes $\xi_{it}$ at some lag $1 \le \ell \le d$ \citep{dahlhaus2000graphical}.

The second network contains undirected edges representing contemporaneous cross-sectional dependence in VAR innovations $\bm\Gamma^{1/2} \bm\vep_t$, denoted by $\mc N^{\undir} = (\mc V, \mc E^{\undir})$.
We have $(i, i') \in \mc E^{\undir}$ if and only if the partial correlation between the $i$-th and $i'$-th elements of $\bm\Gamma^{1/2} \bm\vep_t$ is non-zero, which in turn is given by $- \delta_{ii'}/\sqrt{\delta_{ii} \cdot \delta_{i'i'}}$ where $\bm\Gamma^{-1} = \bm\Delta = [\delta_{ii'}, \, 1 \le i, i' \le p]$ \citep{peng2009partial}.
Hence, the set of edges for $\mc N^{\undir}$ is given by
\begin{align}
\label{eq:net:undir}
\mc E^{\undir} = \l\{ (i, i') \in \mc V \times \mc V: \, i \ne i' \text{ and }
- \frac{\delta_{ii'}}{\sqrt{\delta_{ii} \cdot \delta_{i'i'}}} \ne 0 \r\},
\end{align}

Finally, we can summarise the aforementioned lead-lag and contemporaneous relations between the variables in a single, undirected network $\mc N^{\lr} = (\mc V, \mc E^{\lr})$ by means of the long-run partial correlations of $\bm\xi_t$. 
Let $\bm\Omega = [\omega_{ii'}, \, 1 \le i, i' \le p]$ denote the inverse of the zero-frequency spectral density (a.k.a.\ long-run covariance) of $\bm\xi_t$, which is given by $\bm \Omega = 2\pi \mc A^\top(1) \bm\Delta \mc A(1)$ with $\mc A(z) = \mbf I - \sum_{\ell = 1}^{d} \mbf A_\ell z^\ell$.
Then, the long-run partial correlation between the $i$-th and $i'$-th elements of $\bm\xi_t$, is obtained as $- \omega_{ii'}/\sqrt{\omega_{ii} \cdot \omega_{i'i'}}$ \citep{dahlhaus2000graphical}, so the edge set of $\mc N^{\lr}$ is given by
\begin{align}
\label{eq:net:lr}
\mc E^{\lr} = \l\{ (i, i') \in \mc V \times \mc V: \, i \ne i' \text{ and }
- \frac{\omega_{ii'}}{\sqrt{\omega_{ii} \cdot \omega_{i'i'}}} \ne 0 \r\}.
\end{align}
\subsection{FNETS: Network estimation}    
\label{sec:estimation}

We describe the three-step methodology for estimating the networks $\mc N^{\dir}$, $\mc N^{\undir}$ and $\mc N^{\lr}$.
Throughout, we assume that the number of factors, either $q$ under the more general model in~\eqref{eq:gdfm} or $r$ under the restricted model in~\eqref{eq:static}, and the VAR order $d$ are known, and discuss its selection in \hyperref[sec:tuning]{Tuning parameter selection}. 

\subsubsection{Step~1: Factor adjustment}
\label{sec:step:one}

The autocovariance (ACV) matrices of $\bm\xi_t$, denoted by $\bm\Gamma_\xi(\ell) = \E(\bm\xi_{t - \ell} \bm\xi_t^\top)$ for $\ell \ge 0$ and $\bm\Gamma_\xi(\ell) = (\bm\Gamma_\xi(- \ell))^\top$ for $\ell < 0$, play a key role in network estimation.
Since $\bm\xi_t$ is not directly observed, we propose to adjust for the presence of the factor-driven $\bm\chi_t$ and estimate $\bm\Gamma_\xi(\ell)$.
For this, we adopt a frequency domain-based approach and perform dynamic principal component analysis (PCA).
Spectral density matrix $\bm\Sigma_x(\omega)$ of a time series $\{\mbf X_t\}_{t \in \Z}$ aggregates information of its ACV $\bm\Gamma_x(\ell), \, \ell \in \Z$, at a specific frequency $\omega \in [-\pi, \pi]$, and is obtained by the Fourier transform $\bm\Sigma_x(\omega) = (2\pi)^{-1} \sum_{\ell = -\infty}^\infty \bm\Gamma_x(\ell) \exp(-\iota \ell \omega)$ where $\iota = \sqrt{ - 1}$.
Denoting the sample ACV matrix of $\mbf X_t$ at lag $\ell$ by
\begin{align*}
\wh{\bm\Gamma}_x(\ell) = \frac{1}{n} \sum_{t = \ell + 1}^n \mbf X_{t - \ell} \mbf X_t^\top
\text{ when } \ell \ge 0 \quad \text{and} \quad
\wh{\bm\Gamma}_x(\ell) = (\wh{\bm\Gamma}_x(-\ell))^\top
\text{ when } \ell < 0,
\end{align*}
we estimate the spectral density of $\mbf X_t$ by
\begin{align}
\label{eq:sigma:hat}
\wh{\bm\Sigma}_x(\omega_k) = \frac{1}{2\pi} \sum_{\ell = -m}^m K\l(\frac{\ell}{m}\r)
\wh{\bm\Gamma}_x(\ell) \exp(-\iota \ell \omega_k), 
\end{align} 
where $K(\cdot)$ denotes a kernel, $m$ the kernel bandwidth (for its choice, see \hyperref[sec:tuning]{Tuning parameter selection}) and $\omega_k = 2\pi k / (2m + 1)$ the Fourier frequencies. 
We adopt the Bartlett kernel as $K(\cdot)$ which ensures positive semi-definiteness of $\wh{\bm\Sigma}_x(\omega)$ and also $\wh{\bm\Gamma}_\xi(\ell)$ estimating $\bm\Gamma_\xi(\ell)$ obtained as described below.

Performing PCA on $\wh{\bm\Sigma}_x(\omega_k)$ at each $\omega_k$, 
we obtain the estimator of the spectral density matrix of $\bm\chi_t$ as $\wh{\bm\Sigma}_\chi(\omega_k) = \sum_{j = 1}^q \wh{\mu}_{x, j}(\omega_k) \wh{\mbf e}_{x, j}(\omega_k) (\wh{\mbf e}_{x, j}(\omega_k))^*$, where $\wh{\mu}_{x, j}(\omega_k)$ denotes the $j$-th largest eigenvalue of $\wh{\bm\Sigma}_x(\omega_k)$, $\wh{\mbf e}_{x, j}(\omega_k)$ its associated eigenvector,
and for any vector $\mbf a \in \mathbb{C}^n$, we denote its transposed complex conjugate by $\mbf a^*$. 
Then taking the inverse Fourier transform of $\wh{\bm\Sigma}_\chi(\omega_k), \, -m \le k \le m$, leads to an estimator of $\bm\Gamma_\chi(\ell)$, the ACV matrix of $\bm\chi_t$, as
\begin{align*}
\wh{\bm\Gamma}_\chi(\ell) = \frac{2\pi}{2m + 1} \sum_{k = -m}^m
\wh{\bm\Sigma}_\chi(\omega_k) \exp(\iota \ell \omega_k) \quad \text{for \ } -m \le \ell \le m.
\end{align*}
Finally, we estimate the ACV of $\bm\xi_t$ by
\begin{align}
\label{eq:acv:xi}
\wh{\bm\Gamma}_\xi(\ell) = \wh{\bm\Gamma}_x(\ell) - \wh{\bm\Gamma}_\chi(\ell).
\end{align}

When we assume the restricted factor model in~\eqref{eq:static}, the factor-adjustment step is simplified as it suffices to perform PCA in the time domain, i.e.\ eigenanalysis of the sample covariance matrix $\wh{\bm\Gamma}_x(0)$. Denoting the eigenvector of $\wh{\bm\Gamma}_x(0)$ associated with its $j$-th largest eigenvalue by $\wh{\mbf e}_{x, j}$, we obtain $\wh{\bm\Gamma}_\xi(\ell) = \wh{\bm\Gamma}_x(\ell) - \wh{\mbf E}_x \wh{\mbf E}_x^\top \wh{\bm\Gamma}_x(\ell) \wh{\mbf E}_x \wh{\mbf E}_x^\top$ where $\wh{\mbf E}_x = [\wh{\mbf e}_{x, j}, \, 1 \le j \le r]$.

\subsubsection{Step~2: Estimation of $\mc N^{\dir}$}
\label{sec:step:two}

Recall from~\eqref{eq:net:dir} that $\mc N^{\dir}$ representing Granger causal linkages, has its edge set determined by the VAR transition matrices $\mbf A_\ell, \, 1 \le \ell \le d$. 
By the Yule-Walker equation, we have $\bm\beta = [ \mbf A_1, \dots, \mbf A_{d} ]^\top = {\mbf G(d)}^{-1} \mbf g(d)$, where
\begin{align}
\label{eq:gg}
\mathbf G(d) = \bmx 
\bm\Gamma_\xi(0) & \bm\Gamma_\xi(-1) & \ldots & \bm\Gamma_\xi(-d + 1) 
\\
\bm\Gamma_\xi(1) & \bm\Gamma_\xi(0) & \ldots & \bm\Gamma_\xi(-d + 2)
\\
& & \ddots & 
\\
\bm\Gamma_\xi(d - 1) & \bm\Gamma_\xi(d - 2) & \ldots & \bm\Gamma_\xi(0)
\emx 
\quad \text{and} \quad
\mbf g(d) = \bmx
\bm\Gamma_\xi(1)
\\
\bm\Gamma_\xi(2)
\\
\vdots
\\
\bm\Gamma_\xi(d)
\emx.
\end{align}
We propose to estimate $\bm\beta$ as a regularised Yule-Walker estimator based on $\wh{\mathbf G}(d)$ and $\wh{\mbf g}(d)$, each of which is obtained by replacing $\bm\Gamma_\xi(\ell)$ with $\wh{\bm\Gamma}_\xi(\ell)$ (see~\eqref{eq:acv:xi}) in the definition of $\mbf G(d)$ and $\mbf g(d)$.

For any matrix $\mbf M = [m_{ij}] \in \R^{n_1 \times n_2}$, let $\vert \mbf M \vert_1 = \sum_{i = 1}^{n_1} \sum_{j = 1}^{n_2} \vert m_{ij} \vert$, $\vert \mbf M \vert_\infty = \max_{1 \le i \le n_1} \max_{1 \le j \le n_2} \vert m_{ij} \vert$ and $\tr(\mbf M) = \sum_{i = 1}^{n_1} m_{ii}$ when $n_1 = n_2$.
We consider two estimators of $\bm\beta$.
Firstly, we adopt a Lasso-type estimator which solves an $\ell_1$-regularised $M$-estimation problem
\begin{align}
\label{eq:lasso}
\wh{\bm\beta}^{\las} = \underset{\mbf M \in \Rbb^{pd \times p}}{\arg\min} \
\tr\l( \mbf M^\top \wh{\mathbf G}(d) \mbf M - 2\mbf M^\top\wh{\mbf g}(d) \r)
+ \lambda \vert \mbf M \vert_1
\end{align}
with a tuning parameter $\lambda > 0$.
In the implementation, we solve~\eqref{eq:lasso} via the fast iterative shrinkage-thresholding algorithm (FISTA, \citeauthor{beck2009fast}, \citeyear{beck2009fast}).
Alternatively, we adopt a constrained $\ell_1$-minimisation approach closely related to the Dantzig selector (DS, \citeauthor{candes2007dantzig}, \citeyear{candes2007dantzig}):
\begin{align}
\label{eq:ds}
\wh{\bm\beta}^{\ds} =  \underset{\mbf M \in \Rbb^{pd \times p}}{\arg\min} \ \vert \mbf M \vert_1
\quad \text{subject to} \quad
\l\vert \wh{\mathbf G}(d) \mbf M - \wh{\mbf g}(d) \r\vert_\infty \le \lambda
\end{align}
for some tuning parameter $\lambda > 0$.
We divide~\eqref{eq:ds} into $p$ sub-problems and obtain each column of $\wh{\bm\beta}^{\ds}$ via the simplex algorithm (using the function \code{lp} in \CRANpkg{lpSolve}).

\cite{barigozzi2022fnets} establish the consistency of both $\wh{\bm\beta}^{\las}$ and $\wh{\bm\beta}^{\ds}$ but, as is typically the case for $\ell_1$-regularisation methods, they do not achieve exact recovery of the support of $\bm\beta$.
Hence we propose to estimate the edge set of $\mc N^{\dir}$ by thresholding the elements of $\wh{\bm\beta}$ with some threshold $\mathfrak{t} > 0$, where either $\wh{\bm\beta} = \wh{\bm\beta}^{\las}$ or $\wh{\bm\beta} = \wh{\bm\beta}^{\ds}$, i.e.\
\begin{align}
\label{eq:threshold}
\wt{\bm\beta}(\mathfrak{t}) &= 
\bmx \wh{\beta}_{ij} \cdot \mathbb{I}_{\{\vert \wh{\beta}_{ij} \vert > \mathfrak{t} \}},
\, 1 \le i \le pd, \, 1 \le j \le p \emx.
\end{align}
We discuss cross validation and information criterion methods for selecting $\lambda$, and a data-driven choice of $\mathfrak{t}$, in \hyperref[sec:tuning]{Tuning parameter selection}. 

\subsubsection{Step~3: Estimation of $\mc N^{\undir}$ and $\mc N^{\lr}$}
\label{sec:step:three}

From the definitions of $\mc N^{\undir}$ and $\mc N^{\lr}$ given in~\eqref{eq:net:undir} and~\eqref{eq:net:lr}, their edge sets are obtained by estimating $\bm\Delta = \bm\Gamma^{-1}$ and $\bm\Omega = 2\pi \mc A^\top(1) \bm\Delta \mc A(1)$.
Given $\wh{\bm\beta} = [\wh{\mbf A}_1, \ldots, \wh{\mbf A}_d]^\top$, some estimator of the VAR parameter matrices obtained as in either~\eqref{eq:lasso} or~\eqref{eq:ds}, a natural estimator of $\bm\Gamma$ arises from the Yule-Walker equation
$\bm\Gamma = \bm\Gamma_\xi(0) - \sum_{\ell = 1}^{d} \mbf A_{\ell} \bm\Gamma_\xi(\ell) = \bm\Gamma_\xi(0) - \bm\beta^\top {\mbf g}$,
as $\wh{\bm\Gamma} = \wh{\bm\Gamma}_\xi(0) - \wh{\bm\beta}^\top \wh{\mbf g}$.
In high dimensions, it is not feasible or recommended to directly invert $\wh{\bm\Gamma}$ to estimate $\bm\Delta$.
Therefore, we adopt a constrained $\ell_1$-minimisation method motivated by the CLIME methodology of \cite{cai2011constrained}.

Specifically, the CLIME estimator of $\bm\Delta$ is obtained by first solving
\begin{align}
\label{eq:clime}
\check{\bm\Delta} & = {\arg\min}_{\mbf M \in \Rbb^{p \times p}} \vert \mbf M \vert_1 
\quad \text{subject to} \quad
\l\vert \wh{\bm\Gamma} \mbf M - \mbf I \r\vert_\infty \le \eta,
\end{align}
and applying a symmetrisation step to $\check{\bm\Delta} = [\check\delta_{ii'}, \, 1 \le i, j \le p]$ as
\begin{align} 
\label{eq:delta:clime}
\wh{\bm\Delta} &= [\wh\delta_{ii'}, \, 1 \le i, i' \le p]
\text{ with } 
\wh\delta_{ii'} = \check\delta_{ii'} \cdot \mathbb{I}_{\{\vert \check\delta_{ii'} \vert
\le \vert \check\delta_{i'i} \vert \}}
+ \check\delta_{i'i} \cdot \mathbb{I}_{\{\vert \check\delta_{i'i} \vert
< \vert \check\delta_{ii'} \vert \}}.
\end{align}
for some tuning parameter $\eta > 0$.
\cite{cai2016estimating} propose ACLIME, which improves the CLIME estimator by selecting the parameter $\eta$ in~\eqref{eq:delta:clime} adaptively.
It first produces the estimators of the diagonal entries $\delta_{ii}, 1 \le i \le p$, as in~\eqref{eq:delta:clime} with $\eta_1 = 2 \sqrt{\log (p) / n}$ as the tuning parameter.
Then these estimates are used for adaptive tuning parameter selection in the second step.
We provide the full description of the ACLIME estimator along with the details of its implementation in \hyperref[sec:aclime]{ACLIME estimator} of the Appendix.

Given the estimators $\wh{\mc A}(1) = \mbf I - \sum_{\ell = 1}^d \wh{\mbf A}_\ell$ and $\wh{\bm\Delta}$, we estimate $\bm\Omega$ by $\wh{\bm\Omega} = 2\pi \wh{\mc A}^\top(1) \wh{\bm\Delta} \wh{\mc A}(1)$.
In \cite{barigozzi2022fnets}, $\wh{\bm\Delta}$ and $\wh{\bm\Omega}$ are shown to be consistent in $\ell_\infty$- and $\ell_1$-norms under suitable sparsity assumptions.
However, an additional thresholding step as in~\eqref{eq:threshold} is required to guarantee consistency in estimating the support of $\bm\Delta$ and $\bm\Omega$ and consequently the edge sets of $\mc N^{\undir}$ and $\mc N^{\lr}$.
We discuss data-driven selection of these thresholds and $\eta$ in \hyperref[sec:tuning]{Tuning parameter selection}.

\subsection{FNETS: Forecasting}

Following the estimation procedure, FNETS performs forecasting by estimating the best linear predictor of $\mbf X_{n + a}$ given $\mbf X_t, \, t \le n$, for a fixed integer $a \ge 1$.
This is achieved by separately producing the best linear predictors of $\bm\chi_{n + a}$ and $\bm\xi_{n + a}$ as described below, and then combining them.

\subsubsection{Forecasting the factor-driven component}
\label{sec:common:pred}

For given $a \ge 0$, the best linear predictor of $\bm\chi_{n + a}$ given $\mbf X_t, \, t \le n$, under~\eqref{eq:gdfm} is 
\begin{align*}
\bm\chi_{n + a \vert n} = \sum_{\ell = 0}^\infty \mbf B_{\ell + a} \mbf u_{n  - \ell}.
\end{align*}
\cite{forni2015dynamic} show that the model~\eqref{eq:gdfm} admits a low-rank VAR representation with $\mbf u_t$ as the innovations under mild conditions, and \cite{forni2017dynamic} propose the estimators of $\mbf B_\ell$ and $\mbf u_t$ based on this representation which make use of the estimators of the ACV of $\bm\chi_t$ obtained as described in \hyperref[sec:step:one]{Step 1}.
Then, a natural estimator of $\bm\chi_{n + a \vert n}$ is
\begin{align}
\label{eq:chi:va}
\wh{\bm\chi}^{\va}_{n + a \vert n} = \sum_{\ell = 0}^K \wh{\mbf B}_{\ell + a} \wh{\mbf u}_{n  - \ell}
\end{align}
for some truncation lag $K$.
We refer to $\wh{\bm\chi}^{\va}_{n + a \vert n}$ as the {\it unrestricted} estimator of $\bm\chi_{n + a \vert n}$ as it is obtained without imposing any restrictions on the factor model~\eqref{eq:gdfm}.

When $\bm\chi_t$ admits the static representation in~\eqref{eq:static}, we can show that $\bm\chi_{n + a \vert n} = \bm{\Gamma}_\chi( - a) \mbf E_\chi \bm{\mc M}_\chi^{-1} \mbf E_\chi^\top \bm\chi_n$, where $\bm{\mc M}_\chi \in \R^{r \times r}$ is a diagonal matrix with the $r$ eigenvalues of $\bm\Gamma_\chi(0)$ on its diagonal and $\mbf E_\chi \in \R^{p \times r}$ the matrix of the corresponding eigenvectors; see Section~4.1 of \cite{barigozzi2022fnets} and also \cite{forni2005generalized}.
This suggests an estimator
\begin{align}
\label{eq:chi:static}
\wh{\bm\chi}^{\static}_{n + a \vert n} = \wh{\bm\Gamma}_\chi(-a) \wh{\mbf E}_\chi \wh{\bm{\mc M}}_\chi^{-1} \wh{\mbf E}_\chi^\top \mbf X_n,
\end{align}
where $\wh{\bm{\mc M}}_\chi$ and $\wh{\mbf E}_\chi$ are obtained from the eigendecomposition of $\wh{\bm\Gamma}_\chi(0)$.
We refer to $\wh{\bm\chi}^{\static}_{n + a \vert n}$ as the {\it restricted} estimator of $\bm\chi_{n + a \vert n}$.
As a by-product, we obtain the in-sample estimators of $\bm\chi_t, \, t \le n$, as $\wh{\bm\chi}_{t \vert n} = \wh{\bm\chi}_t$, with either of the two estimators in~\eqref{eq:chi:va} and~\eqref{eq:chi:static}.

\subsubsection{Forecasting the latent VAR process}
\label{sec:idio:pred}

Once the VAR parameters are estimated either as in~\eqref{eq:lasso} or~\eqref{eq:ds}, we produce an estimator of $\bm\xi_{n + a \vert n} = \sum_{\ell = 1}^{d} \mbf A_{\ell} \bm\xi_{n + a - \ell}$, the best linear predictor of $\bm\xi_{n + a}$ given $\mbf X_t, \, t \le n$, as
\begin{align}
\label{eq:idio:best:lin:pred}
\wh{\bm\xi}_{n + a \vert n} = \sum_{\ell = 1}^{\max(1, a) - 1} \wh{\mbf A}_{\ell} \wh{\bm\xi}_{n + a - \ell \vert n} + \sum_{\ell = \max(1, a)}^{d} \wh{\mbf A}_{\ell} \wh{\bm\xi}_{n + a - \ell}.
\end{align}
Here, $\wh{\bm\xi}_{n + 1 - \ell} = \mbf X_{n + 1 - \ell} - \wh{\bm\chi}_{n + 1 - \ell}$ denotes the in-sample estimator of $\bm\xi_{n + 1 - \ell}$, which may be obtained with either of the two (in-sample) estimators of the factor-driven component in~\eqref{eq:chi:va} and~\eqref{eq:chi:static}.

\section{Tuning parameter selection}
\label{sec:tuning}

\subsection{Factor numbers $q$ and $r$}
\label{sec:tuning:factor}

The estimation and forecasting tools of the FNETS methodology require the selection of the number of factors, i.e. $q$ under the unrestricted factor model in~\eqref{eq:gdfm}, and $r$ under the restricted, static factor model in~\eqref{eq:static}.
Under~\eqref{eq:gdfm}, there exists a large gap between the $q$ leading eigenvalues of the spectral density matrix of $\mbf X_t$ and the remainder which diverges with $p$ (see also Figure~\ref{fig:eigen}
We provide two methods for selecting the factor number $q$, which make use of the postulated eigengap using $\wh\mu_{x, j}(\omega_k), \, 1 \le j \le p$, the eigenvalues of the spectral density estimator of $\mbf X_t$ in~\eqref{eq:sigma:hat} at a given Fourier frequency $\omega_k, \, -m \le k \le m$.

\begin{figure}[htb]
\centering
\includegraphics[width = .6\textwidth]{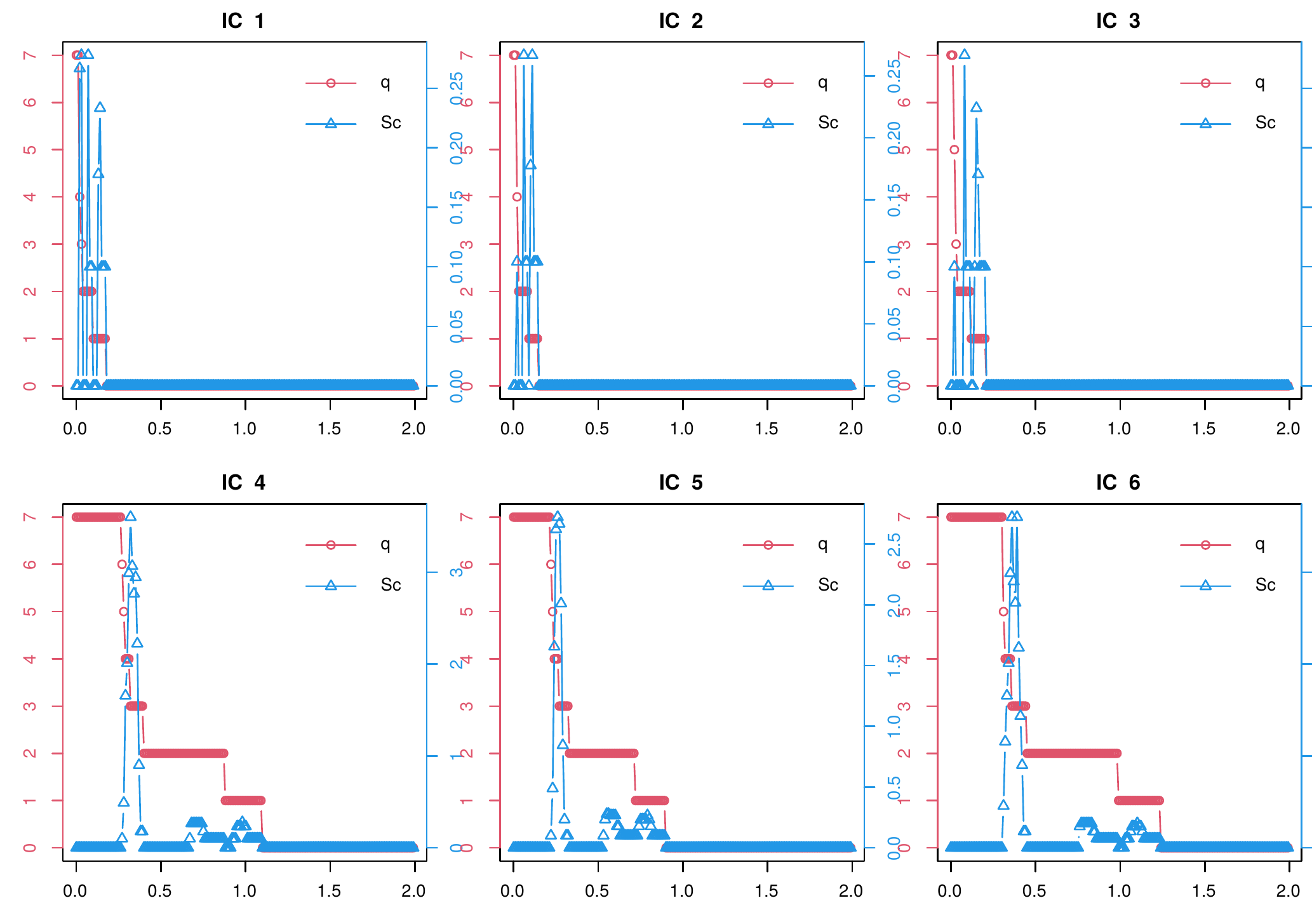}
\caption{Plots of $c$ against $\wh q(n, p, c)$ (in circle, $y$-axis on the left) and $S(c)$ (in triangle, $y$-axis on the right) with the six IC (see \hyperref[sec:factornumber]{Appendix A}) implemented in the function \code{factor.number} of \pkg{fnets}, on a dataset simulated as in \hyperref[sec:package:data]{Data generation} (with $n = 500$, $p = 50$ and $q = 2$).
With the default choice of IC in~\eqref{eq:ic} (IC$_5$), we obtain $\wh q = \wh q(n, p, \wh c) = 2$ correctly estimating $q = 2$.}
\label{fig:qplot}
\end{figure}

\cite{hallin2007determining} propose an information criterion for selecting the number of factors under the model~\eqref{eq:gdfm} and further, a methodology for tuning the multiplicative constant in the penalty.
Define
\begin{align}
\label{eq:ic}
	\text{IC}(b, c) = \log \left(\frac{1}{p} \sum _{j = b + 1}^p \frac{1}{2m + 1} \sum_{k = -m}^m \wh{\mu}_{x, j}(\omega_k) \right) 
 + b \cdot c \cdot \text{pen}(n, p),
\end{align}
where $\text{pen}(n, p) = \min(p, m^2, \sqrt{n / m})^{-1/2}$ by default (for other choices of the information criterion, see \hyperref[sec:factornumber]{Appendix~A}) and $c > 0$ a constant.
Provided that $\text{pen}(n, p) \to 0$ sufficiently slowly, for an arbitrary value of $c$, the factor number $q$ is consistently estimated by the minimiser of $\text{IC}(b, c)$ over $b \in \{0, \ldots, \bar{q}\}$, with some fixed $\bar{q}$ as the maximum allowable number of factors. 
However, this is not the case in finite sample, and \cite{hallin2007determining} propose to simultaneously select~$q$ and~$c$. 
First, we identify $\wh q(n_l, p_l, c) = \arg\min_{0 \le b \le \bar{q}} \text{IC}(n_l, p_l, b, c)$ where $\text{IC}(n_l, p_l, b, c)$ is constructed analogously to $\text{IC}(b, c)$, except that it only involves the sub-sample $\{X_{it}, \, 1 \le i \le p_l, \, 1 \le t \le n_l\}$, for sequences $0 < n_1 < \ldots < n_L = n$ and $0 < p_1 < \ldots < p_L = p$.
Then, denoting the sample variance of $\wh q(n_l, p_l, c), \, 1 \le l \le L$, by $S(c)$, we select $\wh q = \wh q(n, p, \wh c)$ with $\wh c$ corresponding to the second interval of stability with $S(c) = 0$ for the mapping $c \mapsto S(c)$ as $c$ increases from $0$ to some $c_{\max}$ (the first stable interval is where $\bar{q}$ is selected with a very small value of $c$).
Figure~\ref{fig:qplot} plots $\wh q(n, p, c)$ and $S(c)$ for varying values of $c$ obtained from a dataset simulated in \hyperref[sec:package:data]{Data generation}. 
In the implementation of this methodology, we set $n_l = n - (L - l) \lfloor n/20 \rfloor$ and $p_l = \lfloor 3p/4 + lp/40 \rfloor$ with $L = 10$, and $\bar{q} = \min(50, \lfloor \sqrt{\min(n - 1, p)} \rfloor)$.

Alternatively, we can adopt the ratio-based estimator $\wh q = \arg\min_{1 \le b \le \bar{q}} \text{ER}(b)$ proposed in \cite{avarucci2022main}, where 
\begin{align}
\label{eq:er}
\text{ER}(b) = \l( \sum_{k = -m}^m \wh{\mu}_{x, b + 1}(\omega_k) \r)^{-1} \l( \sum_{k = -m}^m \wh{\mu}_{x, b}(\omega_k) \r).
\end{align}

These methods are readily modified to select the number of factors $r$ under the restricted factor model in~\eqref{eq:static}, by replacing $(2m + 1)^{-1} \sum_{k = -m}^m \wh{\mu}_{x, j}(\omega_k)$ with $\wh\mu_{x, j}$, the $j$-th largest eigenvalues of the sample covariance matrix $\wh{\bm\Gamma}_x(0)$.
We refer to \cite{bai2002} and \cite{alessi2010improved} for the discussion of the information criterion-based method in this setting, and \cite{ahn2013eigenvalue} for that of the eigenvalue ratio-based method.

\subsection{Threshold $\mathfrak{t}$}
\label{sec:tuning:thresh}

Motivated by \cite{liu2021simultaneous}, we propose a method for data-driven selection of the threshold $\mathfrak{t}$,
which is applied to the estimators of $\mbf A_\ell, \, 1 \le \ell \le d$, $\bm\Delta$ or $\bm\Omega$ for estimating the edge sets of $\mc N^{\dir}$, $\mc N^{\undir}$ or $\mc N^{\lr}$, respectively; see also~\eqref{eq:threshold}.

\begin{figure}[hbp]
\centering
\includegraphics[width = .4\textwidth]{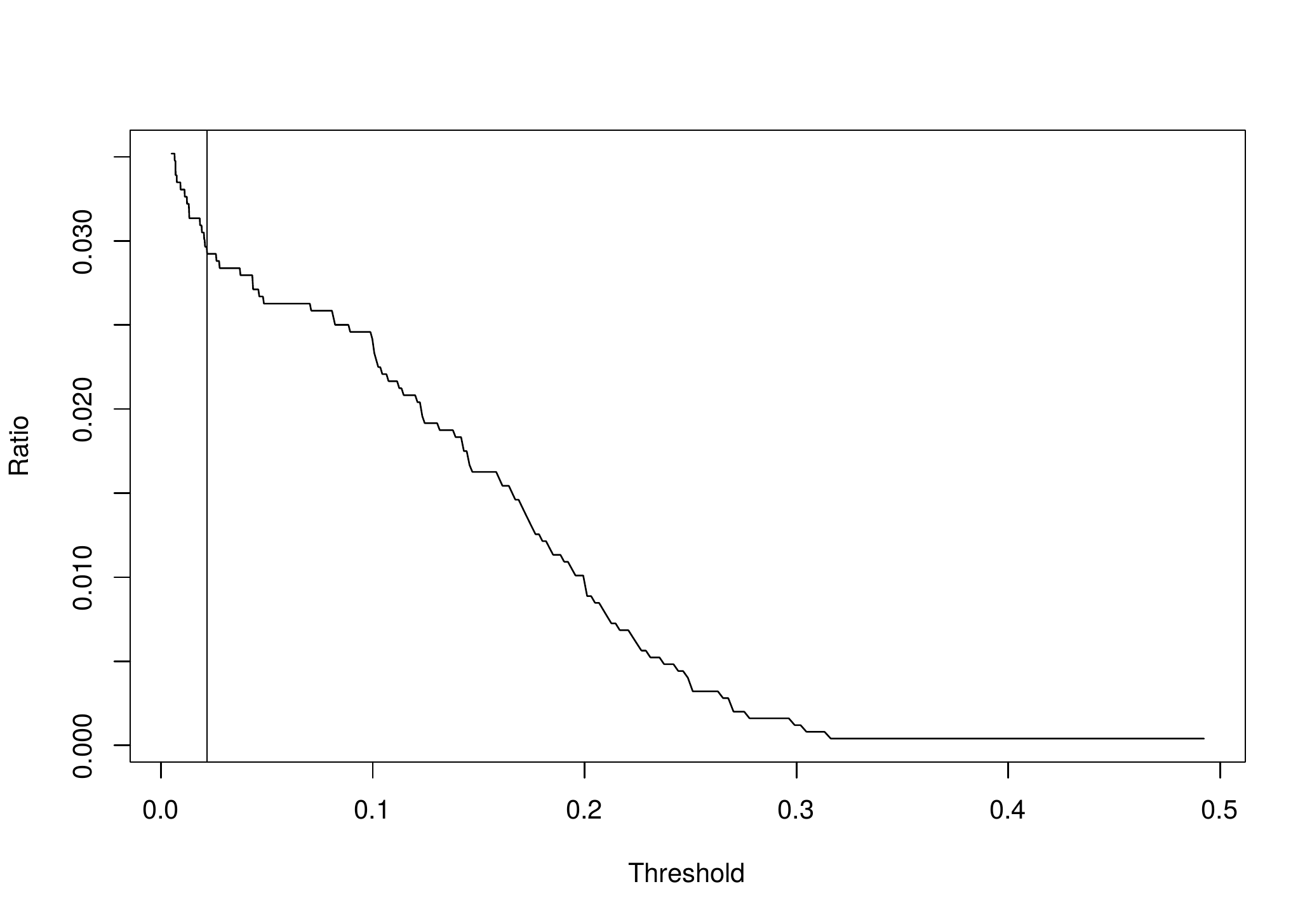}
\includegraphics[width = .4\textwidth]{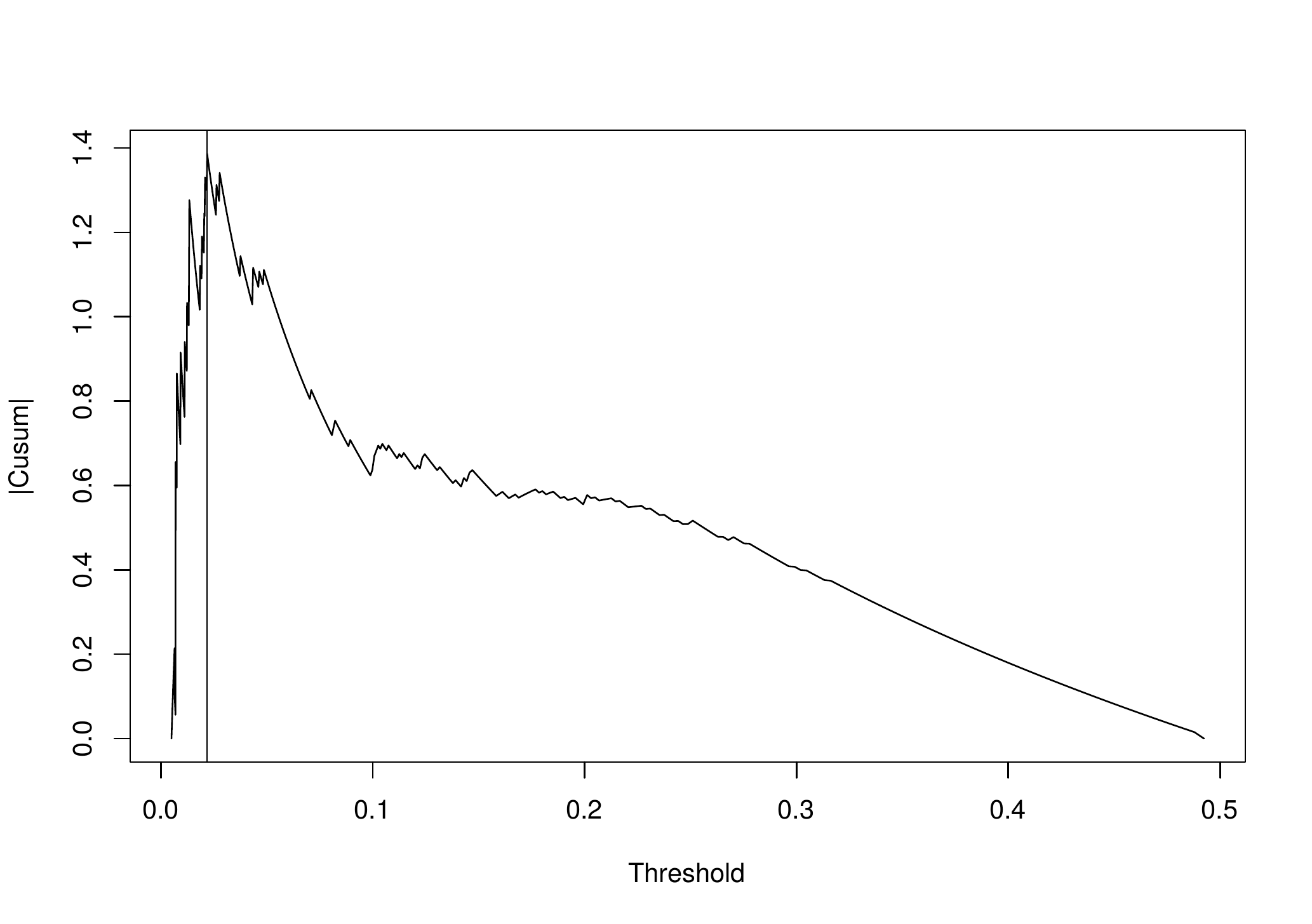}
\caption{Ratio$_k$ (left) and CUSUM$_k$ (right) plotted against $\mathfrak{t}_k$ when $\mbf B = \wh{\bm\beta}^{\las}$ obtained from the data simulated in \hyperref[sec:package:data]{Data generation} with $n = 500$ and $p = 50$, as a Lasso estimator of the VAR parameter matrix, with the selected $\mathfrak{t}_{\text{ada}}$ denoted by the vertical lines.}
\label{fig:thresh}
\end{figure}

Let $\mbf B = [b_{ij}] \in \R^{m \times n}$ denote a matrix for which a threshold is to be selected, i.e.\ $\mbf B$ may be either $\wh{\bm\beta} = [\wh{\mbf A}_1, \ldots, \wh{\mbf A}_d]^\top$, $\wh{\bm\Delta}_0$ ($\wh{\bm\Delta}$ with diagonals set to zero) or $\wh{\bm\Omega}_0$ ($\wh{\bm\Omega}$ with diagonals set to zero) obtained from Steps~2 and~3 of FNETS.
We work with $\wh{\bm\Delta}_0$ and $\wh{\bm\Omega}_0$ since we do not threshold the diagonal entries of $\wh{\bm\Delta}$ and $\wh{\bm\Omega}$.
As such estimators have been shown to achieve consistency in $\ell_\infty$-norm, we expect there exists a large gap between the entries of $\mbf B$ corresponding to true positives and false positives.
Further, it is expected that the number of edges reduces at a faster rate when increasing the threshold from $0$ towards this (unknown) gap, compared to when increasing the threshold from the gap to $\vert \mbf B \vert_\infty$.
Therefore, we propose to identify this gap by casting the problem as that of locating a single change point in the trend of the ratio of edges to non-edges,
\begin{align*}
\text{Ratio}_k = \frac{\vert \mbf B(\mathfrak{t}_k) \vert_0}{ \max( N - \vert \mbf B(\mathfrak{t}_k) \vert_0, 1) }, 
\quad k = 1, \dots, M.
\end{align*}
Here, $\mbf B(\mathfrak{t}) = [b_{ij} \cdot \mathbb{I}_{\{\vert b_{ij} \vert > \mathfrak{t} \}}]$, 
$\vert \mbf B(\mathfrak{t}) \vert_0 = \sum_{i = 1}^{m_1} \sum_{j = 1}^{m_2} \mathbb{I}_{\{\vert b_{ij} \vert > \mathfrak{t} \}}$
and $\{\mathfrak{t}_k, \, 1 \le k \le M: \, 0 = \mathfrak{t}_1 < \mathfrak{t}_2 < \dots < \mathfrak{t}_M = \vert \mbf B \vert_\infty\}$ denotes a sequence of candidate threshold values.  
We recommend using an exponentially growing sequence for $\{\mathfrak{t}_k\}_{k = 1}^M$ since the size of the false positive entries tends to be very small.
The quantity $N$ in the denominator of Ratio$_k$ is set as $N = p^2d$ when $\mbf B = \wh{\bm\beta}$, and $N = p(p - 1)$ when $\mbf B = \wh{\bm\Delta}_0$ or $\mbf B = \wh{\bm\Omega}_0$.
Then, from the difference quotient
\begin{align*} 
\text{Diff}_k = \frac{\text{Ratio}_k - \text{Ratio}_{k - 1}}{\mathfrak{t}_k - \mathfrak{t}_{k - 1}},
\quad k = 2, \ldots, M,
\end{align*}
we compute the cumulative sum (CUSUM) statistic
\begin{align*}
\text{CUSUM}_k = \sqrt{\frac{k (M - k)}{M}} \l\vert \frac{1}{k} \sum_{l = 2}^k \text{Diff}_l - \frac{1}{M - k} \sum_{l = k + 1}^M \text{Diff}_l \r\vert, \quad k = 2, \ldots, M-1,
\end{align*}
and select $\mathfrak{t}_{\text{ada}} = \mathfrak{t}_{k^*}$ with $k^* = {\arg\max}_{2 \le k \le M - 1} \text{CUSUM}_k$.
For illustration, Figure~\ref{fig:thresh} plots Ratio$_k$ and CUSUM$_k$ against candidate thresholds for the dataset simulated in \hyperref[sec:package:data]{Data generation}.

\subsection{VAR order $d$, $\lambda$ and $\eta$}
\label{sec:order:lambda}

\hyperref[sec:step:two]{Step 2} and \hyperref[sec:step:three]{Step 3} of the network estimation methodology of FNETS 
involve the selection of the tuning parameters $\lambda$ and $\eta$ (see~\eqref{eq:lasso}, \eqref{eq:ds} and~\eqref{eq:clime}) and the VAR order $d$.
While there exist a variety of methods available for VAR order selection in fixed dimensions \citep[Chapter~4]{lutk}, the data-driven selection of $d$ in high dimensions remains largely unaddressed with a few exceptions \citep{nicholson2020high, krampe2021, zheng2022interpretable}.
We suggest two methods for jointly selecting $\lambda$ and $d$ for Step~2.
The first method is also applicable for selecting $\eta$ in Step~3.

\subsubsection{Cross validation}
\label{sec:tuning:cv}

Cross validation (CV) methods have popularly been adopted for tuning parameter and model selection.
While some works exist which justify the usage of conventional CV procedure in time series setting in the absence of model mis-specification \citep{bergmeir2018note}, such arguments do not apply to our problem due to the latency of component time series. 
Instead, we propose to adopt a modified CV procedure that bears resemblance to out-of-sample evaluation or rolling forecasting validation \citep{wang2021rate}, for simultaneously selecting $d$ and $\lambda$ in Step~2.
For this, the data is partitioned into $L$ folds, $\mc I_l = \{n^\circ_l + 1, \ldots, n^\circ_{l + 1}\}$ with $n^\circ_l = \min(l \lceil n/L \rceil, n), \, 1 \le l \le L$, and each fold is split into a training set $\mc I^{\text{train}}_l = \{n^\circ_l + 1, \ldots, \lceil (n^\circ_l + n^\circ_{l + 1})/2 \rceil\}$ and a test set $\mc I^{\text{test}}_l = \mc I_l \setminus \mc I^{\text{train}}_l$.
On each fold, $\bm\beta$ is estimated from $\{\mbf X_t, \, t \in \mc I^{\text{train}}_l\}$ as either the Lasso~\eqref{eq:lasso} or the Dantzig selector~\eqref{eq:ds} estimators with $\lambda$ as the tuning parameter and some $b$ as the VAR order, say $\wh{\bm\beta}^{\text{train}}_l(\lambda, b)$, using which we compute the CV measure
\begin{align*}
\text{CV}(\lambda, b) &= \sum_{l = 1}^L \tr\l( \wh{\bm\Gamma}^{\text{test}}_{\xi, l}(0) - (\wh{\bm\beta}_l^{\text{train}}(\lambda, b))^\top \wh{\mbf g}_l^{\text{test}}(b) -  
\r. \\  & \qquad \qquad \qquad \l. 
(\wh{\mbf g}_l^{\text{test}}(b))^\top \wh{\bm\beta}_l^{\text{train}}(\lambda, b)
+  (\wh{\bm\beta}_l^{\text{train}}(\lambda, b))^\top \wh{\mathbf G}_l^{\text{test}}(b)
\wh{\bm\beta}_l^{\text{train}}(\lambda, b) \r),
\end{align*}
where $\wh{\bm\Gamma}_{\xi, l}^{\text{test}}(\ell), \wh{\mathbf G}_l^{\text{test}}(b)$ and $\wh{\mbf g}_l^{\text{test}}(b)$ are generated analogously as $\wh{\bm\Gamma}_\xi(\ell)$, $\wh{\mbf G}(b)$ and $\wh{\mbf g}(b)$, respectively, from the test set $\{\mbf X_t, \, t \in \mc I^{\text{test}}_l\}$.
Although we do not directly observe $\bm\xi_t$, the measure $\text{CV}(\lambda, b)$ gives an approximation of the prediction error.
Then, we select $(\wh{\lambda}, \wh{d}) = \arg\min_{\lambda \in \Lambda, 1 \le b \le \bar{d}} \text{CV} (\lambda, b)$, where $\Lambda$ is a grid of values for $\lambda$, and $\bar{d} \ge 1$ is a pre-determined upper bound on the VAR order.
A similar approach is taken for the selection of $\eta$ with a Burg matrix divergence-based CV measure:
\begin{align*}
\text{CV}(\eta) = \sum_{l = 1}^L \tr\l( \wh{\bm\Delta}_l^{\text{train}}(\eta) \wh{\bm\Gamma}_l^{\text{test}} \r) - \log \l\vert \wh{\bm\Delta}_l^{\text{train}}(\eta) \wh{\bm\Gamma}_l^{\text{test}} \r\vert - p.
\end{align*}
Here, $\wh{\bm\Delta}_l^{\text{train}}(\eta)$ denotes the estimator of $\bm\Delta$ with $\eta$ as the tuning parameter from $\{\mbf X_t, \, t \in \mc I^{\text{train}}_l\}$, and $\wh{\bm\Gamma}_l^{\text{test}}$ the estimator of $\bm\Gamma$ from $\{\mbf X_t, \, t \in \mc I^{\text{test}}_l\}$, see \hyperref[sec:step:three]{Step 3} for the descriptions of the estimators.
In the numerical results reported in \hyperref[sec:sim]{Simulations}, the sample size is relatively small (ranging between $n = 200$ and $n = 500$ while $p \in \{50, 100, 200\}$ and the number of parameters increasing with $p^2$), and we set $L = 1$ which returns reasonably good performance.
When a larger number of observations are available relative to the dimensionality, we may use the number of folds greater than one. 

\subsubsection{Extended Bayesian information criterion}
\label{sec:tuning:ebic}

Alternatively, to select the pair $(\lambda, d)$ in Step~2, we propose to use the extended Bayesian information criterion (eBIC) of \cite{chen2008extended}, originally proposed for variable selection in high-dimensional linear regression.
Let $\wt{\bm\beta}(\lambda, b, \mathfrak{t}_{\text{ada}})$ denote the thresholded version of $\wh{\bm\beta}(\lambda, b)$ as in~\eqref{eq:threshold} with the threshold $\mathfrak{t}_{\text{ada}}$ chosen as described in \hyperref[sec:tuning:thresh]{Threshold $\mathfrak{t}$}.
Then, letting $s(\lambda, b) = \vert \wt{\bm\beta}(\lambda, b, \mathfrak{t}_{\text{ada}}) \vert_0$, we define
\begin{align}
\label{eq:ebic}
\text{eBIC}_{\alpha} (\lambda, b) &= \frac{n}{2} \log \l( \mc L(\lambda, b) \r) +  s(\lambda, b) \log(n) + 2 \alpha  \log{ bp^2 \choose s(\lambda, b)}, \quad \text{where}
\\
\mathcal{L}(\lambda, b) &= \tr\l( \wh{\mathbf G}(b) - (\wt{\bm\beta}(\lambda, b))^\top \wh{\mbf g}(b) - (\wh{\mbf g}(b))^\top \wt{\bm\beta}(\lambda, b) + (\wt{\bm\beta}(\lambda, b))^\top \wh{\mathbf G}(b) \wt{\bm\beta}(\lambda, b) \r).
	\nn
\end{align}
Then, we select $(\wh\lambda, \wh d) = \arg\min_{\lambda \in \Lambda, 1 \le b \le \bar{d}} \text{eBIC}_{\alpha} (\lambda, b)$.
The constant $\alpha \in (0, 1)$ determines the degree of penalisation which may be chosen from the relationship between $n$ and $p$.
Preliminary simulations suggest that $\alpha = 0$ is a suitable choice for the dimensions $(n, p)$ considered in our numerical studies. 

\subsection{Other tuning parameters}

Motivated by theoretical results reported in \cite{barigozzi2022fnets}, we select the kernel bandwidth for Step~1 of FNETS as $m = \lfloor 4 (n/\log(n))^{1/3} \rfloor$.
In forecasting the factor-driven component as in~\eqref{eq:chi:va}, we set the truncation lag at $K = 20$, as it is expected that the elements of $\mbf B_\ell$ decay rapidly as $\ell$ increases for short-memory processes. 

\section{Package overview} 
\label{sec:package}

\pkg{fnets} is available from the Comprehensive R Archive Network (CRAN).
The main function, \code{fnets}, implements the FNETS methodology for the input data and returns an object of S3 class \code{fnets}.
\code{fnets.var} implements Step~2 of the FNETS methodology estimating the VAR parameters only, and is applicable directly for VAR modelling of high-dimensional time series; its outputs are of class \code{fnets}.
\code{fnets.factor.model} performs factor modelling under either of the two models~\eqref{eq:gdfm} and~\eqref{eq:static}, and returns an object of class \code{fm}.
We provide \code{predict} methods for the objects of classes \code{fnets} and \code{fm}, and a \code{plot} method for the objects of the \code{fnets} class. 
We recommend that the input time series for the above functions are to be transformed to stationarity (if necessary) after a unit root test.
In this section, we demonstrate how to use the functions included with the package.

\subsection{Data generation}
\label{sec:package:data}

For illustration, we generate an example dataset of $n = 500$ and $p = 50$ following the model~\eqref{eq:model}.
\pkg{fnets} provides functions for this purpose. 
For given $n$ and $p$, the function \code{sim.var} generates the VAR($1$) process following~\eqref{eq:idio:var} with $d = 1$, $\bm\Gamma$ as supplied to the function ($\bm\Gamma = \mbf I$ by default),
and $\mbf A_1$ generated as described in \hyperref[sec:sim]{Simulations}.
The function \code{sim.unrestricted} generates the factor-driven component under the unrestricted factor model in~\eqref{eq:gdfm} with $q$ dynamic factors ($q = 2$ by default) and the filter $\mc B(L)$ generated as in model~\ref{m:ar} of \hyperref[sec:sim]{Simulations}.
\begin{example}
set.seed(111)
n <- 500
p <- 50
x <- sim.var(n, p)$data + sim.unrestricted(n, p)$data
\end{example} 
Throughout this section, we use the thus-generated dataset in demonstrating \pkg{fnets} unless specified otherwise.
There also exists \code{sim.restricted} which generates the factor-driven component under the restricted factor model in~\eqref{eq:static}.
For all data generation functions, the default is to use the standard normal distribution for generating $\mbf u_t$ and $\bm\vep_t$, while supplying the argument \code{heavy = TRUE}, the innovations are generated from $\sqrt{3/5} \cdot t_5$, the $t$-distribution with $5$ degrees of freedom scaled to have unit variance.
The package also comes attached with pre-generated datasets \code{data.restricted} and \code{data.unrestricted}.

\subsection{Calling \code{fnets} with default parameters}

The function \code{fnets} can be called with the $n \times p$ data matrix \code{x} as the only input, which sets all other arguments to their default choices.
Then, it performs the factor-adjustment under the unrestricted model in~\eqref{eq:gdfm} with $q$ estimated by minimising the IC in~\eqref{eq:ic}.
The VAR parameter matrix is estimated via the Lasso estimator in~\eqref{eq:lasso} with $d = 1$ as the VAR order and the tuning parameters $\lambda$ and $\eta$ chosen via CV, and no thresholding is performed.
This returns an object of class \code{fnets} whose entries are described in Table~\ref{table:output}, and is supported by a \code{print} method as below. 
\begin{example}
fnets(x)

Factor-adjusted vector autoregressive model with 
n: 500, p: 50
Factor-driven common component --------- 
Factor model: unrestricted
Factor number: 2
Factor number selection method: ic
Information criterion: IC5
Idiosyncratic VAR component --------- 
VAR order: 1
VAR estimation method: lasso
Tuning method: cv
Threshold: FALSE
Non-zero entries: 95/2500
Long-run partial correlations --------- 
LRPC: TRUE
\end{example} 

\begin{table}[htb]
\centering
\caption{Entries of S3 objects of class \code{fnets}}
\label{table:output}
{\small
\begin{tabular}{r ll}
\toprule
Name & Description & Type \\
\cmidrule(lr){1-1} \cmidrule(lr){2-2} \cmidrule(lr){3-3}
\code{q}    & Factor number  & integer \\
\code{spec}   & Spectral density matrices for $\mbf X_t$, $\bm\chi_t$ and $\bm\xi_t$ (when \code{fm.restricted = FALSE}) & list \\
\code{acv} & Autocovariance matrices for $\mbf X_t$, $\bm\chi_t$ and $\bm\xi_t$ & list \\
\code{loadings} & Estimates of $\mbf B_\ell, \, 0 \le \ell \le K$ (when \code{fm.restricted = FALSE}) & array\\
& or $\bm\Lambda$ (when \code{fm.restricted = TRUE}) &  \\ 
\code{factors} & Estimates of $\{\mbf u_t\}$ (when \code{fm.restricted = FALSE}) & array \\
& or $\{\mbf F_t\}$ (when \code{fm.restricted = TRUE}) & \\
\code{idio.var} & Estimates of $\mbf A_\ell, \, 1 \le \ell \le d$, and $\bm\Gamma$, and $d$ and $\lambda$ used & list \\
\code{lrpc} & Estimates of $\bm\Delta$, $\bm\Omega$, (long-run) partial correlations and $\eta$ used & list \\
\code{mean.x} & Sample mean vector & vector \\
\code{var.method} & Estimation method for $\mbf A_\ell$ (input parameter) & string \\
\code{do.lrpc} & Whether to estimate the long-run partial correlations (input parameter) & Boolean \\
\code{kern.bw} & Kernel bandwidth (when \code{fm.restricted = FALSE}, input parameter) & double \\ 
\bottomrule
\end{tabular}}
\end{table}

\subsection{Calling \code{fnets} with optional parameters}

We can also specify the arguments of \code{fnets} to control how Steps~1--3 of FNETS are to be performed.
The full model call is as follows:
\begin{example}
out <- fnets(x, center = TRUE, fm.restricted = FALSE, 
  q = c("ic", "er"), ic.op = NULL, kern.bw = NULL,
  common.args = list(factor.var.order = NULL, max.var.order = NULL, trunc.lags = 20, 
  n.perm = 10), var.order = 1, var.method = c("lasso", "ds"),
  var.args = list(n.iter = NULL, n.cores = min(parallel::detectCores() - 1, 3)),
  do.threshold = FALSE, do.lrpc = TRUE, lrpc.adaptive = FALSE,
  tuning.args = list(tuning = c("cv", "bic"), n.folds = 1, penalty = NULL, 
  path.length = 10)
)
\end{example}
Here, we discuss a selection of input arguments.
The \code{center} argument will de-mean the input.
\code{fm.restricted} determines whether to perform the factor-adjustment under the restricted factor model in~\eqref{eq:static} or not.
If the number of factors is known, we can specify \code{q} with a non-negative integer. 
Otherwise, it can be set as \code{"ic"} or \code{"er"} which selects the factor number estimator to be used between~\eqref{eq:ic} and~\eqref{eq:er}.
When \code{q = "ic"}, setting the argument \code{ic.op} as an integer between $1$ and $6$ specifies the choice of the IC (see \hyperref[sec:factornumber]{Appendix A}) where the default is \code{ic.op = 5}. 
\code{kern.bw} takes a positive integer which specifies the bandwidth to be used in Step~1 of FNETS.
The list \code{common.args} specifies arguments for estimating $\mbf B_\ell$ and $\mbf u_t$ under~\eqref{eq:gdfm}, and relates to the low-rank VAR representation of $\bm\chi_t$ under the unrestricted factor model.
\code{var.order} specifies a vector of positive integers to be considered in VAR order selection.
\code{var.method} determines the method for VAR parameter estimation, which can be either \code{"lasso"} (for the estimator in~\eqref{eq:lasso}) or \code{"ds"} (for that in~\eqref{eq:ds}).
The list \code{var.args} takes additional parameters for Step~2 of FNETS, such as the number of gradient descent steps (\code{n.iter}, when \code{var.method = "lasso"}) or the number of cores to use for parallel computing (\code{n.cores}, when \code{var.method = "ds"}).
\code{do.threshold} selects whether to threshold the estimators of $\mbf A_\ell, \, 1 \le \ell \le d$, $\bm\Delta$ and $\bm\Omega$.
It is possible to perform Steps~1--2 of FNETS only 
without estimating $\bm\Delta$ and $\bm\Omega$ by setting \code{do.lrpc = FALSE}.
If \code{do.lrpc = TRUE}, \code{lrpc.adaptive} specifies whether to use the non-adaptive estimator in~\eqref{eq:clime} or the ACLIME estimator.
The list \code{tuning.args} supplies arguments to the CV or eBIC procedures, including the number of folds $L$ (\code{n.folds}), the eBIC parameter $\alpha$ (\code{penalty}, see~\eqref{eq:ebic}) and the length of the grid of values for $\lambda$ and/or $\eta$ (\code{path.length}). 
Finally, it is possible to set only a subset of the arguments of \code{common.args}, \code{var.args} and \code{tuning.args} whereby the unspecified arguments are set to their default values. 

The factor adjustment (Step~1) and VAR parameter estimation (Step~2) functionalities can be accessed individually by calling  \code{fnets.factor.model} and \code{fnets.var}, respectively.
The latter is equivalent to calling \code{fnets} with \code{q = 0} and \code{do.lrpc = FALSE}.
The former returns an object of class \code{fm} which contains the entries of the \code{fnets} object in Table~\ref{table:output} that relate to the factor-driven component only.

\subsection{Network visualisation}
\label{sec:ex:network}

\begin{figure}
\centering
\includegraphics[width = .45\textwidth]{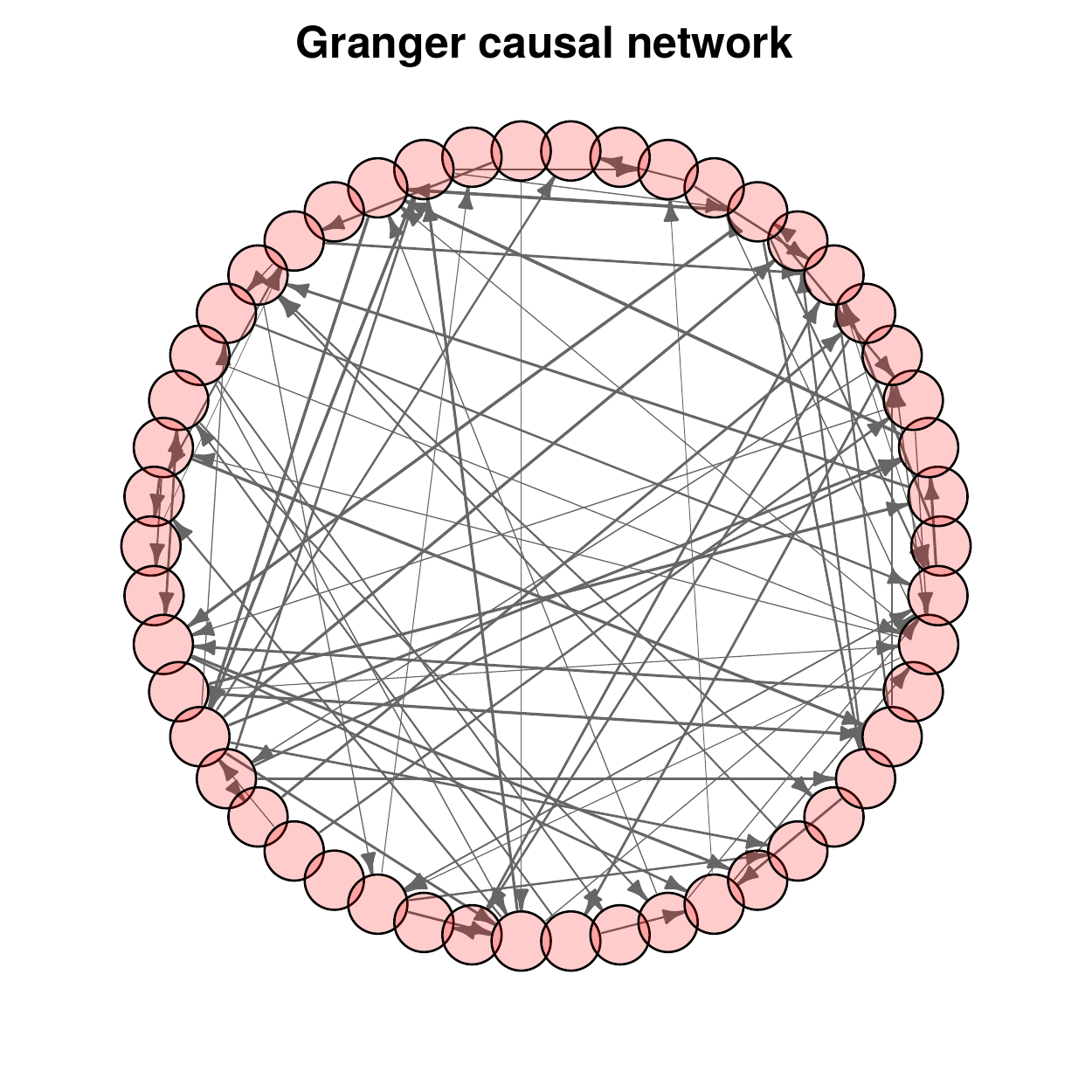}
\includegraphics[width = .45\textwidth]{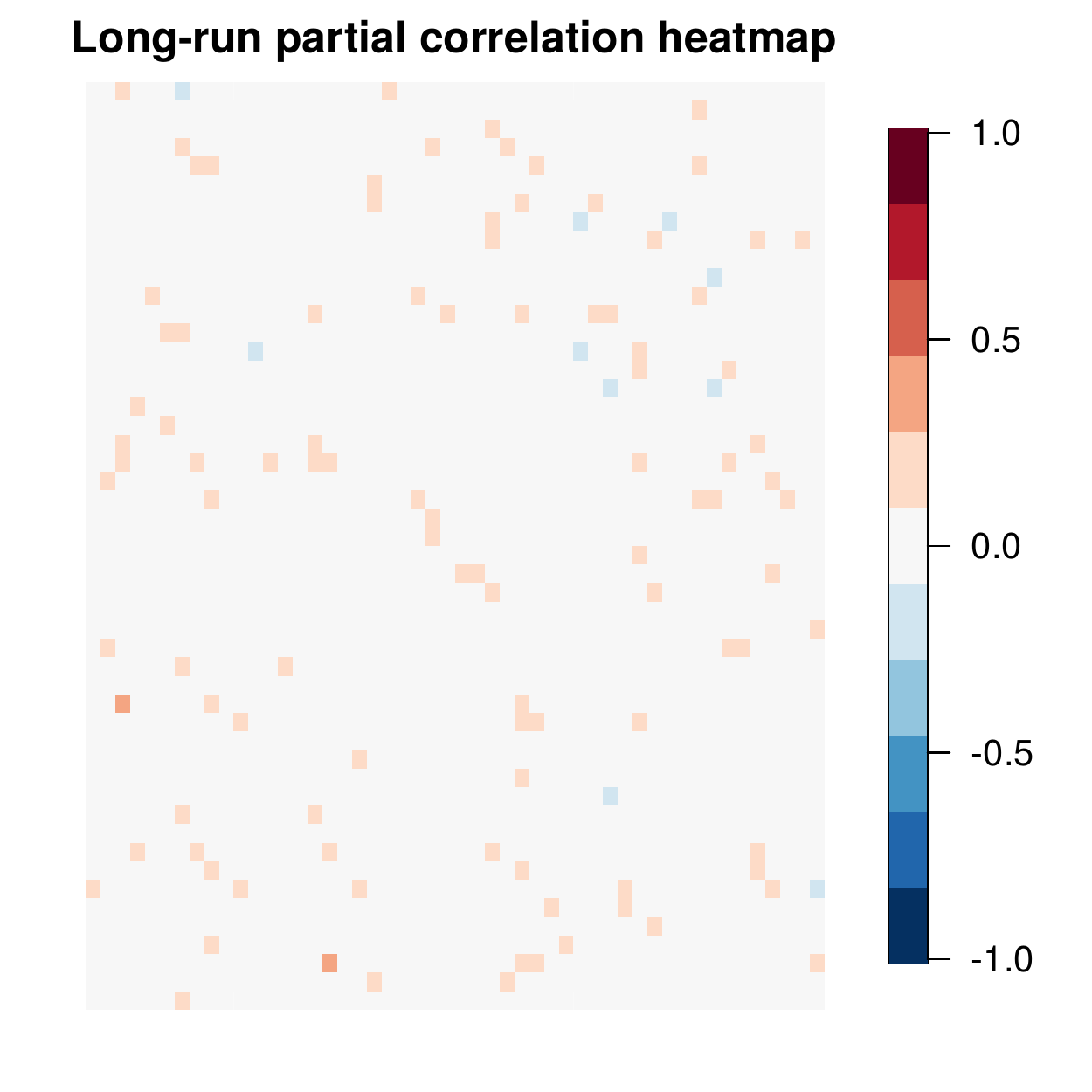}
\caption{Estimated networks for data simulated as in \hyperref[sec:package:data]{Data generation}.
Left: Granger causal network~$\mc N^{\dir}$. A directed arrow from node $i$ to node $i'$ indicates that variable $i$ Granger causes node $i'$, and the edge weights proportional to the size of estimated coefficients are visualised by the edge width.
Right: Long-run partial correlation network $\mc N^{\lr}$ where the edge weights (i.e.\ partial correlations) are visualised by the colour.}
\label{figure:heatmap}
\end{figure}

Using the \code{plot} method available for the objects of class \code{fnets}, we can visualise the Granger network $\mc N^{\dir}$ induced by the estimated VAR parameter matrices, see the left panel of Figure~\ref{figure:heatmap}. 
\begin{example}
plot(out, type = "granger", display = "network")
\end{example}
With \code{display = "network"}, it plots an \code{igraph} object from \CRANpkg{igraph} \citep{igraph}.
Setting the argument \code{type} to \code{"pc"} or \code{"lrpc"}, we can visualise $\mc N^{\undir}$ given by the partial correlations of VAR innovations or $\mc N^{\lr}$ given by the long-run partial correlations of $\bm\xi_t$.
We can instead visualise the networks as a heat map, with the edge weights colour-coded by setting \code{display = "heatmap"}.
We plot $\mc N^{\lr}$ as a heat map in the right panel of Figure~\ref{figure:heatmap} using the following command.
\begin{example}
plot(out, type = "lrpc", display = "heatmap")
\end{example}
It is possible to directly produce an \code{igraph} object from the objects of class \code{fnets} via \code{network} method as: 
\begin{example}
g <- network(out, type = "granger")$network
plot(g, layout = igraph::layout_in_circle(g), 
     vertex.color = grDevices::rainbow(1, alpha = 0.2), vertex.label = NA, 
     main = "Granger causal network")
\end{example}
This produces a plot identical to the left panel of Figure~\ref{figure:heatmap} using the \code{igraph} object \code{g}.

\subsection{Forecasting}

The \code{fnets} objects are supported by the \code{predict} method with which we can forecast the input data \verb+n.ahead+ steps.
For example, we can produce a one-step ahead forecast of $\mbf X_{n + 1}$ as
\begin{example}
pr <- predict(out, n.ahead = 1, fc.restricted = TRUE)
pr$forecast 
\end{example}
The argument \code{fc.restricted} specifies whether to use the estimator $\wh{\bm\chi}^{\static}_{n + h \vert n}$
in~\eqref{eq:chi:static} generated under a restricted factor model~\eqref{eq:static}, or $\wh{\bm\chi}^{\va}_{n + h \vert n}$ in~\eqref{eq:chi:va} generated without such a restriction.
Table~\ref{table:output:predict} lists the entries from the output from \verb+predict.fnets+.
We can similarly produce forecasts from \verb+fnets+ objects output from \verb+fnets.var+, or \verb+fm+ objects from \verb+fnets.factor.model+. 

\begin{table}[htb]
\centering
\caption{Entries of the output from \code{predict.fnets}}
\label{table:output:predict}
{\small
\begin{tabular}{r ll}
\toprule
Name & Description & Type \\
\cmidrule(lr){1-1} \cmidrule(lr){2-2} \cmidrule(lr){3-3} 
\code{forecast}    & $h \times p$ matrix containing the $h$-step ahead forecasts of $\mbf X_t$ & matrix \\
\code{common.predict}    & A list containing & list \\
\code{\$is} & $n \times p$ matrix containing the in-sample estimator of $\bm\chi_t$ & \\ 
\code{\$fc} &  $h \times p$ matrix containing the $h$-step ahead forecasts of $\bm\chi_t$ & \\
\code{\$h} & Input parameter & \\
\code{\$r} & Factor number (only produced when \code{fc.restricted = TRUE}) & \\
\code{idio.predict}    & A list containing \code{is}, \code{fc} and \code{h}, see \code{common.predict}  & list \\
\code{mean.x}    & Sample mean vector  & vector \\
\bottomrule
\end{tabular}}
\end{table}

\subsection{Factor number estimation}

It is of independent interest to estimate the number of factors (if any) in the input dataset. The function \code{factor.number} provides access to the two methods for selecting $q$ described in \hyperref[sec:tuning:factor]{Factor numbers $q$ and~$r$}.
The following code calls the information criterion-based factor number estimation method in~\eqref{eq:ic}, and prints the output: 
\begin{example}
fn <- factor.number(x, fm.restricted = FALSE)
print(fn)

Factor number selection 
Factor model: unrestricted
Method: Information criterion
Number of factors:
IC1: 2
IC2: 2
IC3: 3
IC4: 2
IC5: 2
IC6: 2
\end{example}
Calling \code{plot(fn)} returns Figure~\ref{fig:qplot} which visualises the factor number estimators from six information criteria implemented.
Alternatively, we call the eigenvalue ratio-based method in~\eqref{eq:er} as
\begin{example}
fn <- factor.number(x, method = "er", fm.restricted = FALSE)
\end{example}
In this case, \code{plot(fn)} produces a plot of $\text{ER}(b)$ against the candidate factor number $b \in \{1, \ldots, \bar{q}\}$.

\subsection{Visualisation of tuning parameter selection procedures}
\label{sec:package:order}

\begin{figure}[htb]
\centering
 \includegraphics[width = .8\textwidth]{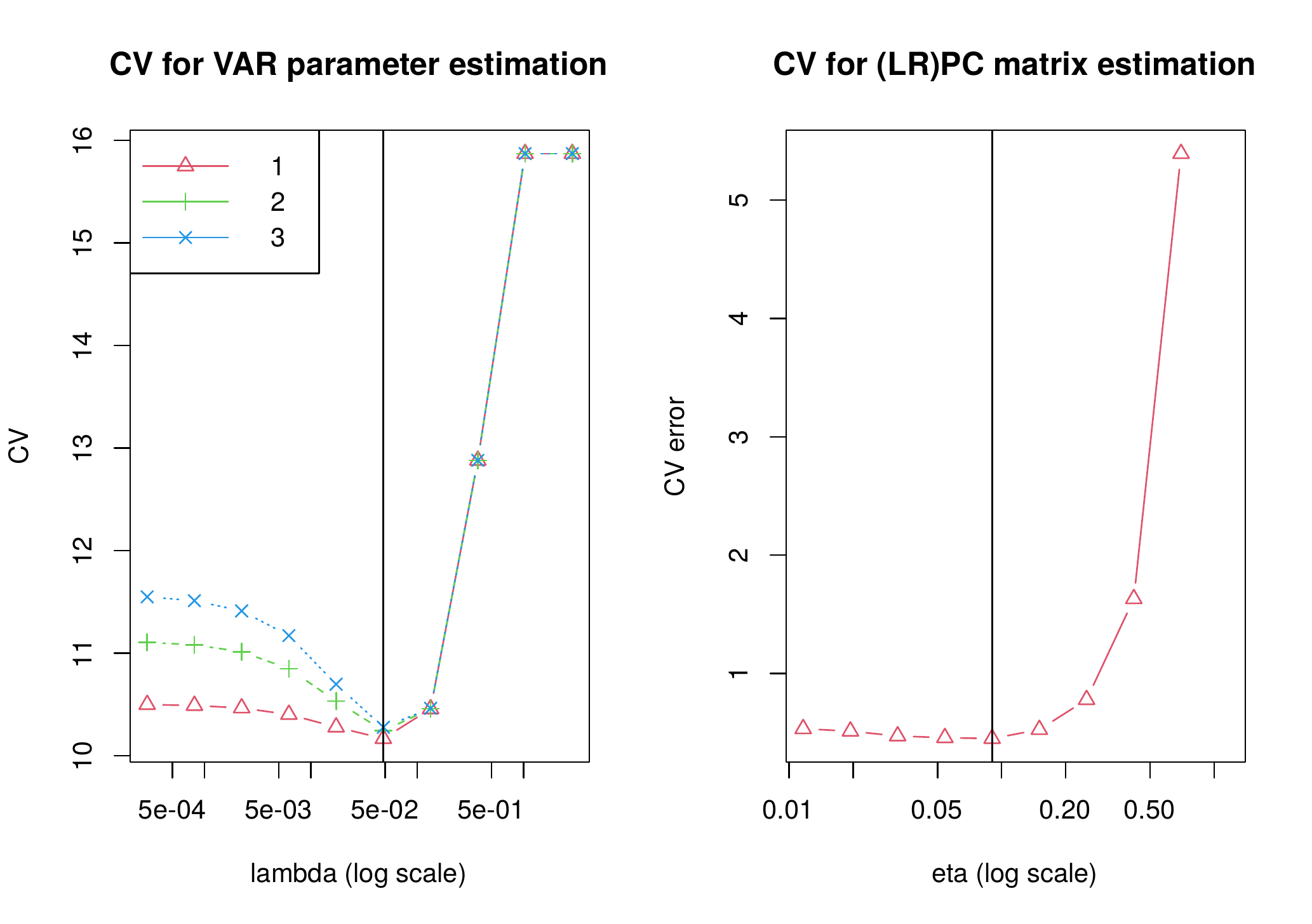} 
\caption{
Plots of CV$(\lambda, b)$ against $\lambda$ with $b \in \{1, 2, 3\}$ (left) and CV$(\eta)$ against $\eta$ (right). Vertical lines denote where the minimum CV measure is attained with respect to $\lambda$ and $\eta$, respectively.}
\label{figure:order_ex}
\end{figure}

The method for threshold selection discussed in \hyperref[sec:tuning:thresh]{Threshold $\mathfrak{t}$} 
is implemented by the \code{threshold} function, which returns objects of \code{threshold} class supported by \code{print} and \code{plot} methods.
\begin{example}
th <- threshold(out$idio.var$beta)
th

Thresholded matrix 
Threshold: 0.0297308643
Non-zero entries: 62/2500
\end{example}
The call \code{plot(th)} generates Figure~\ref{fig:thresh}.
Additionally, we provide tools for visualising the tuning parameter selection results adopted in Steps~2 and~3 of FNETS (see \hyperref[sec:order:lambda]{VAR order $d$, $\lambda$ and $\eta$}). 
These tools are accessible from both \code{fnets} and \code{fnets.var} by calling the \code{plot} method with the argument \code{display = "tuning"}, e.g.\ 
\begin{example}
set.seed(111)
n <- 500
p <- 10
x <- sim.var(n, p)$data
out1 <- fnets(x, q = 0, var.order = 1:3, tuning.args = list(tuning = "cv"))
plot(out1, display = "tuning")        
\end{example}
This generates the two plots reported in Figure~\ref{figure:order_ex} which visualise the CV errors computed as described in \hyperref[sec:tuning:cv]{Cross validation} and, in particular, the left plot shows that the VAR order is correctly selected by this approach.
When \code{tuning.args} contains \code{tuning =  "bic"}, the results from the eBIC method described in \hyperref[sec:tuning:ebic]{Extended Bayesian information criterion} adopted in Step~2, is similarly visualised in place of the left panel of Figure~\ref{figure:order_ex}.

\section{Simulations}
\label{sec:sim}

\cite{barigozzi2022fnets} provide comprehensive simulation results on the estimation and forecasting performance of FNETS in comparison with competing methodologies.
Therefore in this paper, we focus on assessing the performance of the methods for selecting tuning parameters such as the threshold and VAR order discussed in \hyperref[sec:tuning]{Tuning parameter selection}.
Additionally in \hyperref[sec:sim:adaptive]{Appendix~B}, we compare the adaptive and the non-adaptive estimators in estimating $\bm\Delta$ and also investigate how their performance is carried over to estimating $\bm\Omega$.

\subsection{Settings}

We consider the following data generating processes for the factor-driven component $\bm\chi_t$:
\begin{enumerate}[label = (C\arabic*)]
\item \label{m:ar} Taken from \cite{forni2017dynamic},
$\chi_{it}$ is generated as a sum of AR processes
$\chi_{it} = \sum_{j = 1}^q a_{ij} (1 - \alpha_{ij} L)^{-1} u_{jt}$ with $q = 2$,
where $u_{jt} \sim_{\iid} \mc N(0, 1)$, $a_{ij} \sim_{\iid} \mc U[-1, 1]$ and $\alpha_{ij} \sim_{\iid} \mc U[-0.8, 0.8]$ with $\mc U[a, b]$ denoting a uniform distribution.
Then, $\bm\chi_t$ does not admit a static representation in~\eqref{eq:static}.

\item \label{m:oracle} $\bm\chi_t = \mbf 0$, i.e.\ the VAR process is directly observed as $\mbf X_t = \bm\xi_t$.
\end{enumerate}

For generating a VAR($d$) process $\bm\xi_t$, we first generate a directed Erd\H{o}s-R\'{e}nyi random graph $\mc N = (\mc V, \mc E)$ on $\mc V = \{1, \ldots, p\}$ with the link probability $1/p$, and set entries of $\mbf A_d$ such that $A_{d, ii'} = 0.275$ when $(i, i') \in \mc E$ and $A_{d, ii'} = 0$ otherwise. 
Also, we set $\mbf A_\ell = \mbf O$ for $\ell < d$.
The VAR innovations are generated as below.
\begin{enumerate}[label = (E\arabic*)]
\item \label{e:one} Gaussian with the covariance matrix $\bm\Gamma = \bm\Delta^{-1} = \mbf I$.

\item \label{e:four} Gaussian with the covariance matrix $\bm\Gamma = \bm\Delta^{-1}$ such that $\delta_{ii} = 1$, $\delta_{i, i + 1} = \delta_{i + 1, i} = 0.6$, $\delta_{i, i + 2} = \delta_{i + 2, i} = 0.3$, and $\delta_{ii'} = 0$ for $\vert i - i' \vert \ge 3$.
\end{enumerate}
For each setting, we generate $100$ realisations.

\subsection{Results: Threshold selection}

\begin{figure}[htb!]
\centering
\includegraphics[width = .8\textwidth]{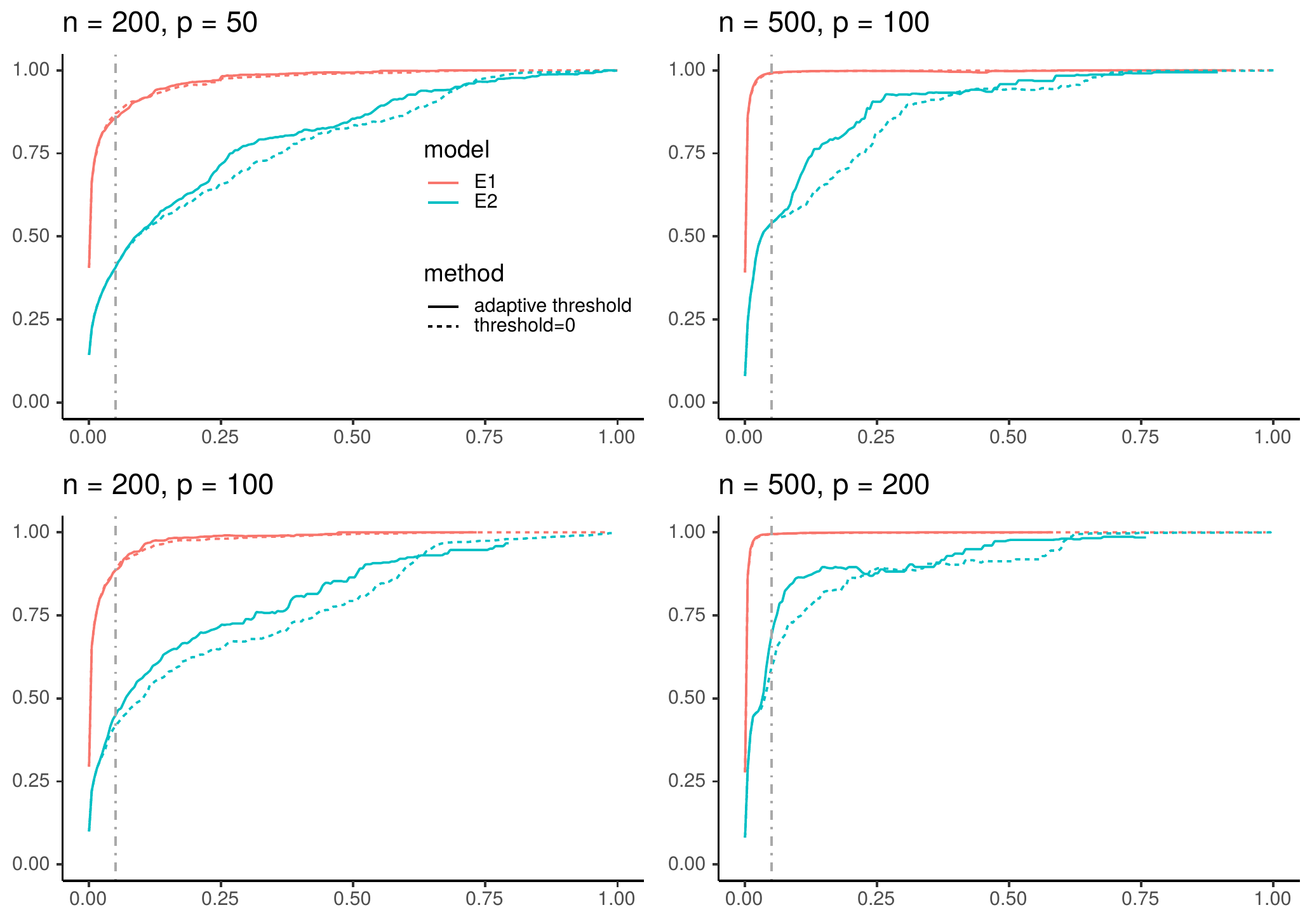}
\caption{\small ROC curves of TPR against FPR for $\wt{\bm\beta}(\mathfrak{t})$~\eqref{eq:threshold} (with $\wh{\bm\beta} = \wh{\bm\beta}^{\las}$) when $\mathfrak{t} =\mathfrak{t}_{ada}$ and $\mathfrak{t} = 0$ in recovering the support of $\bm\Omega$, averaged over $100$ realisations.
Vertical lines indicate FPR $= 0.05$ }
\label{fig:sim:omegathresh}
\end{figure}

We assess the performance of the adaptive threshold.
We generate $\bm\chi_t$ as in~\ref{m:ar} and fix $d = 1$ for generating $\bm\xi_t$ and further, treat $d$ as known.
We consider $(n, p) \in \{(200, 50), (200, 100), (500, 100), (500, 200)\}$. 
Then we estimate $\bm\Omega$ using the thresholded Lasso estimator of $\mbf A_1$ (see~\eqref{eq:lasso} and~\eqref{eq:threshold}) with two choices of thresholds, $\mathfrak{t} = \mathfrak{t}_{\text{ada}}$ generated as described in \hyperref[sec:tuning:thresh]{Threshold $\mathfrak{t}$} and $\mathfrak{t} = 0$.
To assess the performance of $\wh{\bm\Omega} = [\wh \omega_{ii'}]$ in recovering of the support of $\bm\Omega = [\omega_{ii'}]$, i.e.\ $\{(i, i'): \, \omega_{ii'} \ne 0 \}$, we plot the receiver operating characteristic (ROC) curves of true positive rate (TPR) against false positive rate (FPR), where
\begin{align*}
\text{TPR} = \frac{ \vert \{ (i, i'): \, \wh \omega_{ii'} \ne 0 \text{ and } \omega_{ii'} \ne 0 \} \vert }{\vert \{ (i, i'): \, \omega_{ii'} \ne 0 \} \vert}
\quad \text{and} \quad
\text{FPR} = \frac{ \vert \{ (i, i'): \, \wh \omega_{ii'} \ne 0 \text{ and } \omega_{ii'} = 0 \} \vert }{\vert \{ (i, i'): \, \omega_{ii'} = 0 \} \vert}.
\end{align*}
Figure~\ref{fig:sim:omegathresh} plots the ROC curves averaged over $100$ realisations when $\mathfrak{t} = \mathfrak{t}_{\text{ada}}$ and $\mathfrak{t} = 0$.
When $\bm\Delta = \mbf I$ under~\ref{e:one}, we see little improvement from adopting $\mathfrak{t}_{\text{ada}}$ as the support recovery performance is already good even without thresholding.
However, when $\bm\Delta \ne \mbf I$ under~\ref{e:four}, the adaptive threshold leads to improved support recovery especially when the sample size is large. 
Tables~\ref{table:thresholdbeta}~and~\ref{table:thresholdomega} in \hyperref[sec:appendix:sim]{Appendix C} additionally report the errors in estimating $\mbf A_1$ and $\bm\Omega$ with and without thresholding, where we see little change is brought by thresholding.
In summary, we conclude that the estimators already perform reasonably well without thresholding, and the adaptive threshold $\mathfrak{t}_{\text{ada}}$ brings marginal improvement in support recovery which is of interest in network estimation.

\subsection{Results: VAR order selection}
\label{sec:sim:order}
 
We compare the performance of the CV and eBIC methods proposed in \hyperref[sec:order:lambda]{VAR order $d, \lambda$ and $\eta$} for selecting the order of the VAR process.
Here, we consider the case when $\bm\chi_t = \mbf 0$ (setting~\ref{m:oracle}) and when $\bm\xi_t$ is generated under~\ref{e:one} with $d \in \{1, 3\}$. 
We set $(n, p) \in \{(200, 10), (200, 20), (500, 10), (500, 20)\}$ where the range of $p$ is in line with the simulation studies conducted in the relevant literature (see e.g.\ \cite{zheng2022interpretable}).
We consider $\{1, 2, 3, 4\}$ as the candidate VAR orders.
Figure~\ref{fig:order} and Table~\ref{table:order} in \hyperref[sec:appendix:sim]{Appendix C} show that CV works reasonably well regardless of $d \in \{1, 3\}$, with slightly better performance observed together with the DS estimator.
On the other hand, eBIC tends to over-estimate the VAR order when $d = 1$ while under-estimating it when $d = 3$, and hence is less reliable compared to the CV method. 

\begin{figure}[htb!]
\centering
\includegraphics[width = .8\textwidth]{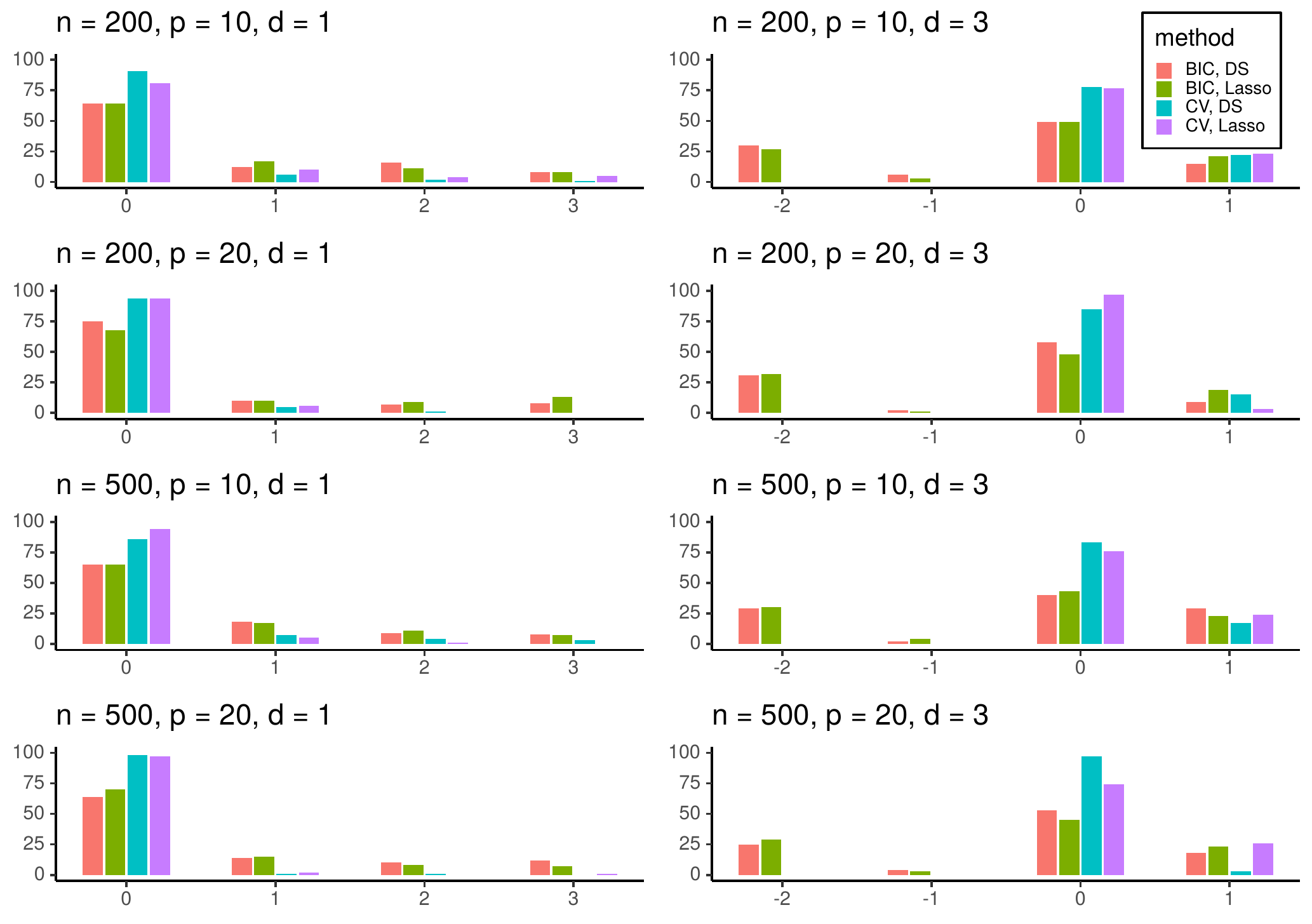}
\caption{Box plots of $\wh{d} - d$ over $100$ realisations when the VAR order is selected by the CV and eBIC methods in combination with the Lasso~\eqref{eq:lasso} and the DS~\eqref{eq:ds} estimators.}
\label{fig:order}
\end{figure}

\section{Data example}
\label{sec:real}

\subsection{Energy price data}
\label{sec:real:energy}

Electricity is more difficult to store than physical commodities which results in high volatility and seasonality in spot prices \citep{han2022extremal}.
Global market deregulation has increased the volume of electricity trading, which promotes the development of better forecasting and risk management methods. 
We analyse a dataset of node-specific prices in the PJM (Pennsylvania, New Jersey and Maryland) power pool area in the United States, accessed using \url{dataminer2.pjm.com}.
There are four node types in the panel, which are Zone, Aggregate, Hub and Extra High Voltage (EHV); 
for their definitions, see Table~\ref{table:definitions} and for the names and types of $p = 50$ nodes, see Table~\ref{table:data:info}, all found in \hyperref[sec:real:data]{Appendix D}.
The series we model is the sum of the real time congestion price and marginal loss price or, equivalently, the difference between the spot price at a given location and the overall system price, where the latter can be thought of as an observed factor in the local spot price.
These are obtained as hourly prices and then averaged over each day as per \cite{maciejowska2013forecasting}.
We remove any short-term seasonality by subtracting a separate mean for each day of the week. Since the energy prices may take negative values, we adopt the inverse hyperbolic sine transformation as in \cite{uniejewski2017variance} for variance stabilisation. 

\subsection{Network estimation}
\label{sec:real:network}
 
We analyse the data collected between 01/01/2021 and 19/07/2021 ($n = 200$).
The information criterion in~\eqref{eq:ic} returns a single factor ($\wh q = 1$), and $\wh d =1 $ is selected by CV. 
See Figure~\ref{fig:real:lasso} for the heat maps visualising the three networks $\mc N^{\dir}$, $\mc N^{\undir}$ and $\mc N^{\lr}$ described in \hyperref[sec:networks]{Networks}, which are produced by \pkg{fnets}.

\begin{widefigure}[htbp!]
\centering
\begin{tabular}{ccc}
\includegraphics[width = .3\textwidth]{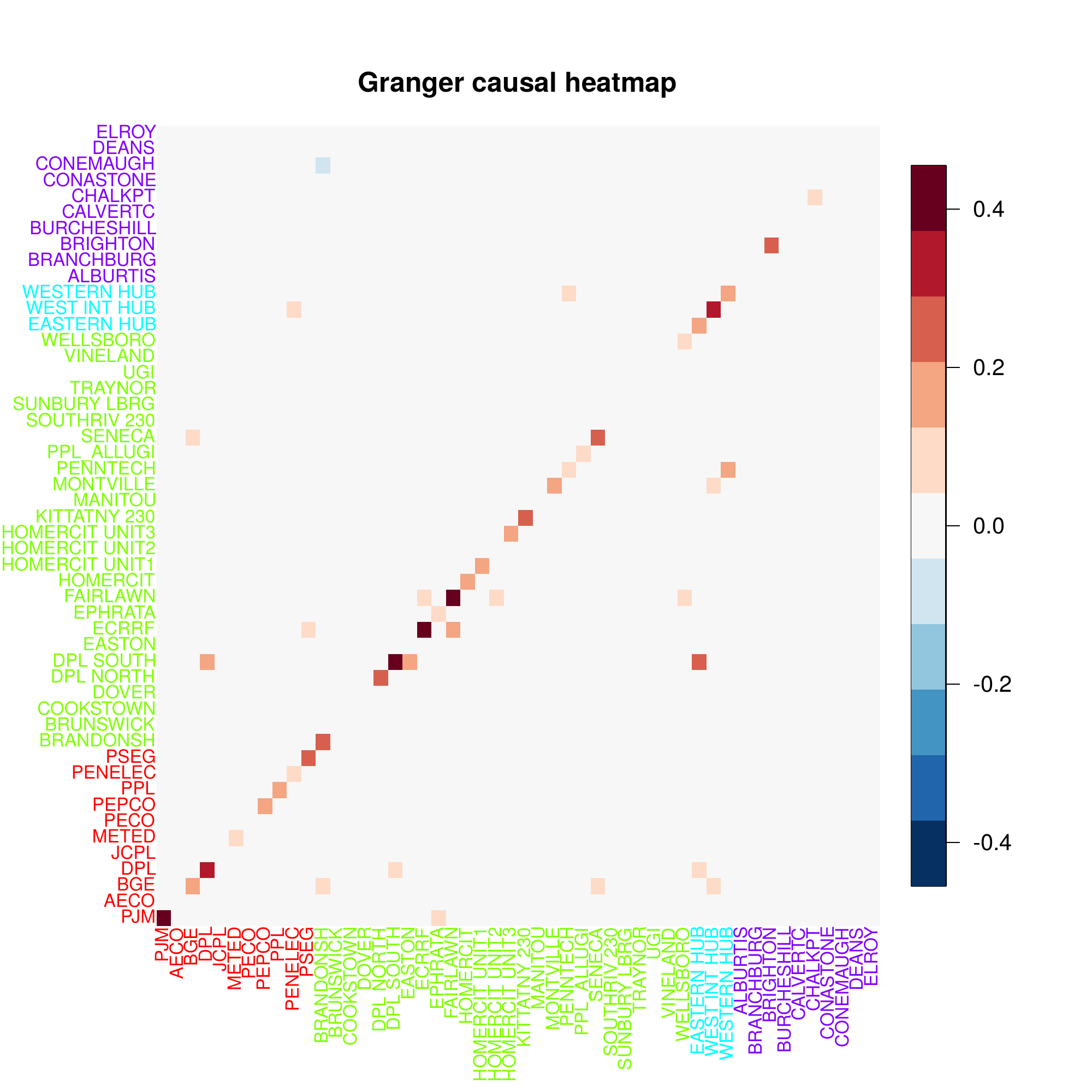} 
&\includegraphics[width = .3\textwidth]{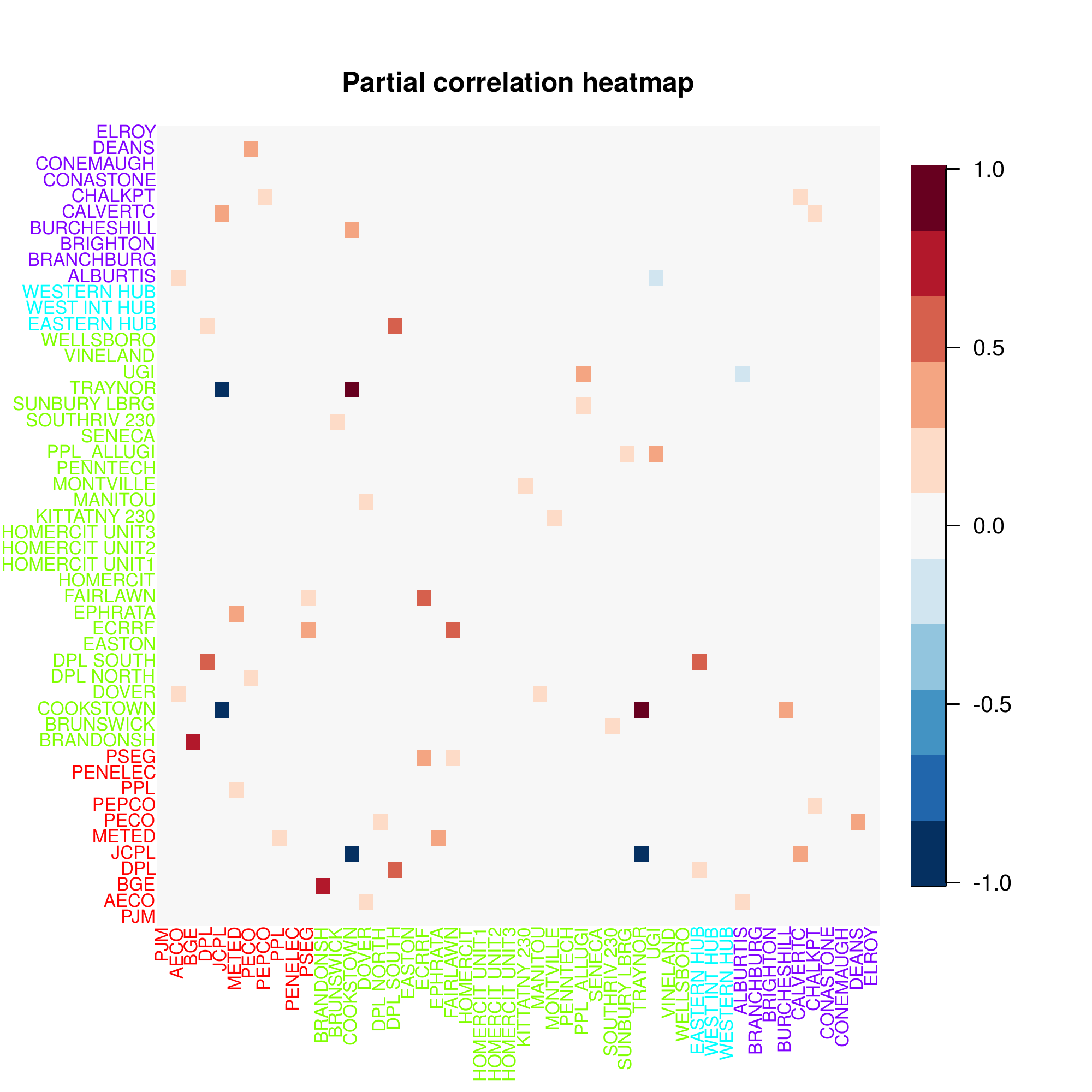} 
&\includegraphics[width = .3\textwidth]{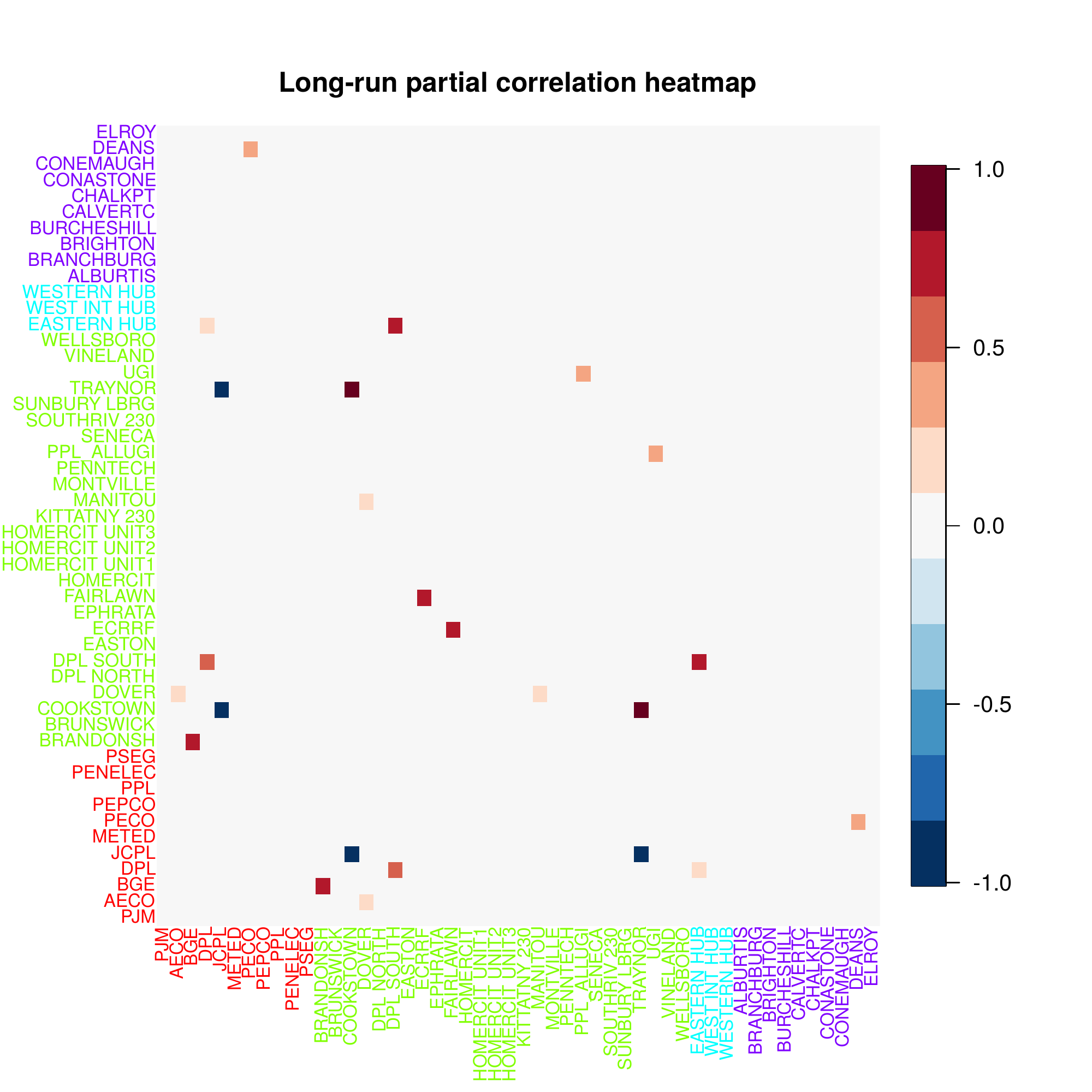} 
\end{tabular}
\caption{Heat maps of the three networks underlying the energy price data collected over the period 01/01/2021--19/07/2021. 
Left: $\mc N^{\dir}$ obtained with the Lasso estimator~\eqref{eq:lasso} combined with the adaptive threshold $\mathfrak{t}_{\text{ada}}$. Middle: $\mc N^{\undir}$ obtained with the ACLIME estimator of $\bm\Delta$. Right: $\mc N^{\lr}$ obtained by combining the estimators of VAR parameters and $\bm\Delta$.
In the axis labels, Zone-type nodes are coloured in red, 
Aggregate-types in green, 
Hub-types in blue
and EHV-types in purple.}
\label{fig:real:lasso}
\end{widefigure}

The non-zero entries of the VAR parameter matrix estimates tend to take positive values, indicating that high energy prices are persistent and spill over to other nodes.
Considering the node types, Hub-type nodes (blue) tend to have out-going edges to nodes of different types, which reflects the behaviour of the electrical transmission system.
Some Zone-type nodes (red) have several in-coming edges from Aggregate-types (green) and Hub-types, while EHV-types (purple) have few edges in $\mc N^{\dir}$, which carries forward to $\mc N^{\lr}$ where we observe that those Zone-type nodes have strong long-run correlations with other nodes while EHV-types do not.
 
\section{Summary}
\label{sec:summary}

We introduce the R package \pkg{fnets} which implements the FNETS methodology proposed by \cite{barigozzi2022fnets} for network estimation and forecasting of high-dimensional time series exhibiting strong correlations. 
It further implements data-driven methods for selecting tuning parameters, and provides tools for high-dimensional time series factor modelling under the GDFM which are of independent interest. 
The efficacy of our package is demonstrated on both
real and simulated datasets.

\bibliography{fnets}

\address{Dom Owens\\
  School of Mathematics, University of Bristol\\
  Supported by EPSRC Centre for Doctoral Training (EP/S023569/1)\\ 
  \email{dom.owens@bristol.ac.uk}}
  
\address{Haeran Cho\\
  School of Mathematics, University of Bristol\\
  Supported by the Leverhulme Trust (RPG-2019-390)\\
  \email{haeran.cho@bristol.ac.uk}}

\address{Matteo Barigozzi\\
  Department of Economics, Universit\`a di Bologna\\
  Supported by MIUR (PRIN 2017, Grant 2017TA7TYC)\\
  \email{matteo.barigozzi@unibo.it}}

\clearpage

\appendix

\section{Appendix A: Information criteria for factor number selection}
\label{sec:factornumber}

Here we list information criteria for factor number estimation which are implemented in \pkg{fnets} and accessible by the functions \code{fnets}, \code{fnets.factor.model} and \code{factor.number} by setting the argument \code{ic.op} at an integer belonging to $\{1, \ldots, 6\}$.
When \code{fm.restricted = FALSE}, we have
\begin{itemize}
    \item[IC$_1$:] 	$ \left(\frac{1}{p} \sum _{j = b + 1}^p \frac{1}{2m + 1} \sum_{k = -m}^m \wh{\mu}_{x, j}(\omega_k) \right)
 + b \cdot c \cdot( m^{-2} + \sqrt{m / n} + p^{-1}) \cdot \log(\min (p, m^2, \sqrt{n / m} ))$,
     \item[IC$_2$:] 	$ \left(\frac{1}{p} \sum _{j = b + 1}^p \frac{1}{2m + 1} \sum_{k = -m}^m \wh{\mu}_{x, j}(\omega_k) \right)
 + b \cdot c \cdot (\min(p, m^2, \sqrt{n / m}))^{-1/2}$, 
      \item[IC$_3$:] 	$ \left(\frac{1}{p} \sum _{j = b + 1}^p \frac{1}{2m + 1} \sum_{k = -m}^m \wh{\mu}_{x, j}(\omega_k) \right) 
 + b \cdot c \cdot (\min(p, m^2, \sqrt{n / m}))^{-1} \cdot \log(\min(p, m^2, \sqrt{n / m}))$,
    \item[IC$_4$:] 	$\log \left(\frac{1}{p} \sum _{j = b + 1}^p \frac{1}{2m + 1} \sum_{k = -m}^m \wh{\mu}_{x, j}(\omega_k) \right) 
 + b \cdot c \cdot( m^{-2} + \sqrt{m / n} + p^{-1}) \cdot \log(\min (p, m^2, \sqrt{n / m} ))$,
     \item[IC$_5$:] 	$\log \left(\frac{1}{p} \sum _{j = b + 1}^p \frac{1}{2m + 1} \sum_{k = -m}^m \wh{\mu}_{x, j}(\omega_k) \right) 
 + b \cdot c \cdot (\min(p, m^2, \sqrt{n / m}))^{-1/2}$,
      \item[IC$_6$:] 	$\log \left(\frac{1}{p} \sum _{j = b + 1}^p \frac{1}{2m + 1} \sum_{k = -m}^m \wh{\mu}_{x, j}(\omega_k) \right) 
 + b \cdot c \cdot (\min(p, m^2, \sqrt{n / m}))^{-1} \cdot \log(\min(p, m^2, \sqrt{n / m}))$ .
\end{itemize}
When \code{fm.restricted = TRUE}, we use one of 
\begin{itemize}
    \item[IC$_1$:] $ \left(\frac{1}{p} \sum _{j = b + 1}^p \wh{\mu}_{x, j} \right)
 + b \cdot c \cdot (n + p) / (n p) \cdot \log(n p / (n + p))$,
     \item[IC$_2$:] $ \left(\frac{1}{p} \sum _{j = b + 1}^p \wh{\mu}_{x, j} \right)
 + b \cdot c \cdot (n + p) / (n p) \cdot \log(n p / (n + p))$,
      \item[IC$_3$:] 	$ \left(\frac{1}{p} \sum _{j = b + 1}^p \wh{\mu}_{x, j} \right) 
 + b \cdot c \cdot \log(\min(n, p)) / (\min(n, p))$,
    \item[IC$_4$:] $ \log\left(\frac{1}{p} \sum _{j = b + 1}^p \wh{\mu}_{x, j} \right)
 + b \cdot c \cdot (n + p) / (n p) \cdot \log(n p / (n + p))$,
     \item[IC$_5$:] $ \log\left(\frac{1}{p} \sum _{j = b + 1}^p \wh{\mu}_{x, j} \right)
 + b \cdot c \cdot (n + p) / (n p) \cdot \log(n p / (n + p))$,
      \item[IC$_6$:] 	$ \log\left(\frac{1}{p} \sum _{j = b + 1}^p \wh{\mu}_{x, j} \right) 
 + b \cdot c \cdot \log(\min(n, p)) / (\min(n, p))$.
\end{itemize}
Whether \code{fm.restricted = FALSE} or not, the default choice is \code{ic.op = 5}.
      
\section{Appendix B: ACLIME estimator}
\label{sec:aclime}

We provide a detailed description of the adaptive extension of the CLIME estimator of $\bm\Delta$ in~\eqref{eq:clime}, extending the methodology proposed in \cite{cai2016estimating} for precision matrix estimation in the independent setting.
Let $\wh{\bm\Gamma}^* = \wh{\bm\Gamma} + n^{-1}{\mbf I}$ and $\eta_1 = 2 \sqrt{{\log (p)} / {n}}$ .

\begin{enumerate}[label = Step~\arabic*:]
\setlength\itemsep{0em}
\item Let $\check{\bm\Delta}^{(1)} = [\check{\delta}_{ii'}^{(1)}]$ be the solution to
\begin{align}
\label{eq:adap:step1}
& \check{\bm\Delta}^{(1)}_{\cdot i'}  = {\arg\min}_{\mbf m \in \R^p} \vert \mbf m \vert_1 
\quad \text{subject to} 
\\
& \l\vert ( \wh{\bm\Gamma}^* \mbf m - \mbf e_{i'})_i \r\vert \le \eta_1 (\wh{\gamma}_{ii} \vee \wh{\gamma}_{i'i'})  m_{i'} \ \forall \ 1 \le i \le p \text{ \ and \ }  m_{i'} > 0, \nn
\end{align}
for $i' = 1, \ldots, p$. 
Then we obtain truncated estimates 
\begin{align*}
	\wh{\delta}_{ii}^{(1)} = \check{\delta}_{ii}^{(1)} \cdot  \mathbb{I}_{\{\vert \wh{\gamma}_{ii} \vert \le \sqrt{n/\log(p)} \}} + 
	 \sqrt{\frac{\log(p)}{n}} \cdot \mathbb{I}_{\{\vert \wh{\gamma}_{ii} \vert > \sqrt{n/\log(p)} \}}.
\end{align*}

\item We obtain 
\begin{align*}
\check{\bm\Delta}^{(2)}_{\cdot i'} & = {\arg\min}_{\mbf m \in \Rbb^p } \vert \mbf m \vert_1 
\quad \text{subject to} \quad 
\l\vert ( \wh{\bm\Gamma}^* \mbf m - \mbf e_{i'})_i \r\vert \le \eta_2 \sqrt{\wh{\gamma}_{ii} \wh{\delta}_{i'i'}^{(1)} } \quad \forall \ 1 \le i \le p,
\end{align*}
where $\eta_2 > 0$ is a tuning parameter.
Since $\check{\bm\Delta}^{(2)}$ is not guaranteed to be symmetric, 
the final estimator is obtained after a symmetrisation step:
\begin{align}
\label{eq:delta:aclime}
\wh{\bm\Delta}_{ada} &= [\wh\delta_{ii'}, \, 1 \le i, i' \le p]
\text{ with } 
\wh\delta_{ii'}^{(2)} = \check\delta_{ii'}^{(2)} \cdot \mathbb{I}_{\{\vert \check\delta_{ii'}^{(2)} \vert
\le \vert \check\delta_{i'i}^{(2)} \vert \}}
+ \check\delta_{i'i}^{(2)} \cdot \mathbb{I}_{\{\vert \check\delta_{i'i}^{(2)} \vert
< \vert \check\delta_{ii'}^{(2)} \vert \}}.
\end{align}
\end{enumerate}

The constraints in \eqref{eq:adap:step1} incorporate the parameter in the right-hand side.
To use linear programming software to solve this, we formulate the constraints for each $1\le i' \le p$ as
\begin{align*}
\forall 1\le i\le p, \quad ( (\wh{\bm\Gamma}^* - \bm Q^{i'}) \mbf m - \mbf e_{i'})_i \le   0,\\
\forall 1\le i\le p, \quad -( (\wh{\bm\Gamma}^* + \bm Q^{i'}) \mbf m - \mbf e_{i'})_i \le 0,\\
  m_{i'} > 0.
\end{align*}
where $\bm Q^{i'}$ has entries $q_{ii'} = \eta_1 (\wh{\gamma}_{ii} \vee \wh{\gamma}_{i'i'})$ in column $i'$ and $0$ elsewhere.

\section{Appendix C: Additional simulation results}
\label{sec:appendix:sim}

\subsection{Threshold selection}

Tables~\ref{table:thresholdbeta}~and~\ref{table:thresholdomega} report the errors in estimating $\mbf A_1$ and $\bm\Omega$ when the threshold $\mathfrak{t} = \mathfrak{t}_{\text{ada}}$ or $\mathfrak{t} = 0$ is applied to the estimator of $\mbf A_1$ obtained by either the Lasso~\eqref{eq:lasso} or the DS~\eqref{eq:ds} estimators. 
With a matrix $\bm\gamma$ as an estimand
we measure the estimation error of its estimator $\wh{\bm\gamma}$
using the following (scaled) matrix norms:
\begin{align*}
L_F = \frac{\Vert \wh{\bm\gamma} - \bm\gamma \Vert_F}{\Vert \bm\gamma \Vert_F} \quad \text{and} \quad L_2 = \frac{\Vert \wh{\bm\gamma} - \bm\gamma \Vert}{\Vert \bm\gamma \Vert}.
\end{align*}

\begin{table}[htb!]
\caption{Errors in estimating $\mbf A_1$ with $\mathfrak{t} \in \{0, \mathfrak{t}_{\text{ada}}\}$ in combination with the Lasso~\eqref{eq:lasso} and the DS~\eqref{eq:ds} estimators, measured by $L_F$ and $L_2$, averaged over $100$ realisations (with standard errors reported in brackets). 
We also report the average TPR when FPR $= 0.05$ and the corresponding standard error.
See \hyperref[sec:sim:order]{Results: Threshold selection} in the main text for further information.} 
\label{table:thresholdbeta}
\centering
\resizebox{\columnwidth}{!}
{\scriptsize 
\begin{tabular}{ccc cccccc cccccc}
\toprule 
 &  &  & \multicolumn{6}{c}{$\mathfrak{t} = 0$} &  \multicolumn{6}{c}{$\mathfrak{t}=\mathfrak{t}_{\text{ada}}$} \\
 \cmidrule(lr){4-9}\cmidrule(lr){10-15}
 &  &  & \multicolumn{3}{c}{$\wh{\bm\beta}^{\las}$} & \multicolumn{3}{c}{$\wh{\bm\beta}^{\ds}$} & \multicolumn{3}{c}{$\wh{\bm\beta}^{\las}$} & \multicolumn{3}{c}{$\wh{\bm\beta}^{\ds}$}   \\
 \cmidrule(lr){4-6}\cmidrule(lr){7-9} \cmidrule(lr){10-12} \cmidrule(lr){13-15}
Model & n & p & TPR & $L_F$ & $L_2$ & TPR & $L_F$ & $L_2$ & TPR & $L_F$ & $L_2$ & TPR & $L_F$ & $L_2$ \\
\cmidrule(lr){1-3}\cmidrule(lr){4-9} \cmidrule(lr){10-15}
\ref{e:one} & 200 & 50 & 0.9681 & 0.6234 & 0.7204 & 0.8991 & 0.4299 & 0.3747 & 0.9413 & 0.6226 & 0.7204 & 0.6932 & 0.4487 & 0.3960 \\
 &  &  & (0.050) & (0.081) & (0.118) & (0.096) & (0.280) & (0.225) & (0.112) & (0.088) & (0.121) & (0.216) & (0.256) & (0.206) \\
 & 200 & 100 & 0.9398 & 0.6696 & 0.8113 & 0.8810 & 0.5772 & 0.4362 & 0.8832 & 0.6710 & 0.8132 & 0.6491 & 0.6025 & 0.4642 \\
 &  &  & (0.091) & (0.096) & (0.096) & (0.094) & (0.449) & (0.271) & (0.182) & (0.108) & (0.100) & (0.246) & (0.418) & (0.250) \\
 & 500 & 100 & 0.9990 & 0.4648 & 0.6682 & 0.9304 & 0.2740 & 0.2604 & 0.9971 & 0.4608 & 0.6645 & 0.7237 & 0.2806 & 0.2699 \\
 &  &  & (0.003) & (0.054) & (0.094) & (0.065) & (0.158) & (0.138) & (0.010) & (0.056) & (0.095) & (0.199) & (0.133) & (0.111) \\
 & 500 & 200 & 0.9986 & 0.5068 & 0.7729 & 0.9167 & 0.3680 & 0.3882 & 0.9964 & 0.5023 & 0.7637 & 0.7095 & 0.3889 & 0.4014 \\
 &  &  & (0.003) & (0.058) & (0.081) & (0.076) & (0.196) & (0.134) & (0.006) & (0.061) & (0.082) & (0.256) & (0.187) & (0.126) \\ \midrule
\ref{e:four} & 200 & 50 & 0.9595 & 0.6375 & 0.7075 & 0.8828 & 0.4673 & 0.4280 & 0.9442 & 0.6356 & 0.7079 & 0.6720 & 0.4835 & 0.4433 \\
 &  &  & (0.053) & (0.077) & (0.094) & (0.107) & (0.324) & (0.255) & (0.064) & (0.079) & (0.096) & (0.212) & (0.303) & (0.241) \\
 & 200 & 100 & 0.9624 & 0.6200 & 0.6909 & 0.8093 & 0.4519 & 0.4090 & 0.9435 & 0.6175 & 0.6913 & 0.5903 & 0.4765 & 0.4324 \\
 &  &  & (0.072) & (0.079) & (0.089) & (0.100) & (0.385) & (0.251) & (0.093) & (0.082) & (0.090) & (0.182) & (0.371) & (0.243) \\
 & 500 & 100 & 0.9970 & 0.4657 & 0.5533 & 0.9304 & 0.3434 & 0.3621 & 0.9958 & 0.4638 & 0.5525 & 0.8384 & 0.3370 & 0.3634 \\
 &  &  & (0.006) & (0.056) & (0.076) & (0.089) & (0.158) & (0.153) & (0.008) & (0.058) & (0.077) & (0.182) & (0.140) & (0.144) \\
 & 500 & 200 & 0.9981 & 0.4702 & 0.5658 & 0.9205 & 0.3684 & 0.3740 & 0.9945 & 0.4686 & 0.5665 & 0.8154 & 0.3663 & 0.3803 \\
 &  &  & (0.003) & (0.065) & (0.091) & (0.088) & (0.182) & (0.162) & (0.014) & (0.068) & (0.093) & (0.205) & (0.159) & (0.145) \\ \bottomrule
\end{tabular} }
\end{table}

\begin{table}[htb!]
\caption{Errors in estimating $\bm\Omega$ with $\mathfrak{t} \in \{0, \mathfrak{t}_{\text{ada}}\}$ applied to the estimator of $\mbf A_1$ in combination with the Lasso~\eqref{eq:lasso} and the DS~\eqref{eq:ds} estimators, measured by $L_F$ and $L_2$, averaged over $100$ realisations (with standard errors reported in brackets). 
We also report the average TPR when FPR $= 0.05$ and the corresponding standard error.
See \hyperref[sec:sim:order]{Results: Threshold selection} in the main text for further information.} 
\label{table:thresholdomega}
\centering
\resizebox{\columnwidth}{!}
{\scriptsize 
\begin{tabular}{ccc cccccc cccccc}
\toprule 
 &  &  & \multicolumn{6}{c}{$\mathfrak{t} = 0$} &  \multicolumn{6}{c}{$\mathfrak{t}=\mathfrak{t}_{\text{ada}}$} \\
 \cmidrule(lr){4-9}\cmidrule(lr){10-15}
 &  &  & \multicolumn{3}{c}{$\wh{\bm\beta}^{\las}$} & \multicolumn{3}{c}{$\wh{\bm\beta}^{\ds}$} & \multicolumn{3}{c}{$\wh{\bm\beta}^{\las}$} & \multicolumn{3}{c}{$\wh{\bm\beta}^{\ds}$}   \\
 \cmidrule(lr){4-6}\cmidrule(lr){7-9} \cmidrule(lr){10-12} \cmidrule(lr){13-15}
Model & n & p & TPR & $L_F$ & $L_2$ & TPR & $L_F$ & $L_2$ & TPR & $L_F$ & $L_2$ & TPR & $L_F$ & $L_2$ \\
\cmidrule(lr){1-3}\cmidrule(lr){4-9} \cmidrule(lr){10-15}
\ref{e:one} & 200 & 50 & 0.8714 & 0.4143 & 0.5553 & 0.8622 & 0.4217 & 0.5691 & 0.8685 & 0.4145 & 0.5559 & 0.8640 & 0.4217 & 0.5695 \\
 &  &  & (0.108) & (0.048) & (0.066) & (0.119) & (0.054) & (0.070) & (0.118) & (0.049) & (0.067) & (0.121) & (0.055) & (0.070) \\
 & 200 & 100 & 0.8827 & 0.4320 & 0.5890 & 0.8961 & 0.4379 & 0.5949 & 0.8684 & 0.4326 & 0.5892 & 0.8867 & 0.4386 & 0.5960 \\
 &  &  & (0.084) & (0.050) & (0.072) & (0.080) & (0.046) & (0.065) & (0.139) & (0.052) & (0.074) & (0.120) & (0.048) & (0.066) \\
 & 500 & 100 & 0.9909 & 0.3311 & 0.4916 & 0.9886 & 0.3391 & 0.4989 & 0.9928 & 0.3303 & 0.4901 & 0.9901 & 0.3380 & 0.4975 \\
 &  &  & (0.016) & (0.031) & (0.069) & (0.021) & (0.036) & (0.065) & (0.015) & (0.032) & (0.069) & (0.018) & (0.037) & (0.066) \\
 & 500 & 200 & 0.9942 & 0.3520 & 0.5287 & 0.9916 & 0.3511 & 0.5400 & 0.9954 & 0.3512 & 0.5273 & 0.9672 & 0.3528 & 0.5399 \\
 &  &  & (0.009) & (0.038) & (0.054) & (0.018) & (0.045) & (0.065) & (0.008) & (0.039) & (0.055) & (0.129) & (0.055) & (0.072) \\ \midrule
\ref{e:four} & 200 & 50 & 0.4074 & 0.7831 & 0.8353 & 0.4027 & 0.7942 & 0.8335 & 0.4063 & 0.7832 & 0.8353 & 0.4045 & 0.7943 & 0.8336 \\
 &  &  & (0.073) & (0.089) & (0.072) & (0.087) & (0.079) & (0.034) & (0.072) & (0.089) & (0.072) & (0.089) & (0.079) & (0.034) \\
 & 200 & 100 & 0.4178 & 0.8406 & 0.8690 & 0.3541 & 0.9119 & 0.8879 & 0.4486 & 0.8407 & 0.8690 & 0.4038 & 0.9120 & 0.8880 \\
 &  &  & (0.091) & (0.108) & (0.036) & (0.107) & (0.126) & (0.045) & (0.091) & (0.108) & (0.036) & (0.123) & (0.126) & (0.045) \\
 & 500 & 100 & 0.5405 & 0.8267 & 0.8118 & 0.5632 & 0.7910 & 0.7953 & 0.5406 & 0.8267 & 0.8117 & 0.5628 & 0.7910 & 0.7951 \\
 &  &  & (0.111) & (0.125) & (0.047) & (0.122) & (0.166) & (0.062) & (0.111) & (0.125) & (0.047) & (0.123) & (0.166) & (0.062) \\
 & 500 & 200 & 0.5951 & 0.8713 & 0.8519 & 0.6487 & 0.8184 & 0.8259 & 0.6918 & 0.8713 & 0.8519 & 0.7101 & 0.8184 & 0.8258 \\
 &  &  & (0.175) & (0.165) & (0.088) & (0.159) & (0.182) & (0.090) & (0.148) & (0.165) & (0.088) & (0.122) & (0.182) & (0.090) \\ \bottomrule
\end{tabular} }
\end{table}

\subsection{VAR order selection}

Table~\ref{table:order} reports the results of VAR order estimation over $100$ realisations.

\begin{table}[htb!]
\caption{Distribution of $\wh{d} - d$ over $100$ realisations when the VAR order is selected by the CV and eBIC methods in combination with the Lasso~\eqref{eq:lasso} and the DS~\eqref{eq:ds} estimators, see \hyperref[sec:sim:order]{Results: VAR order selection} in the main text for further information.}
\label{table:order}
\centering
\resizebox{\columnwidth}{!}
{\scriptsize 
\begin{tabular}{ccc cccc cccc cccc cccc}
\toprule
&	&	&	\multicolumn{8}{c}{CV} &						\multicolumn{8}{c}{eBIC} 						\\	\cmidrule(lr){4-11} \cmidrule(lr){12-19}
&	&	& \multicolumn{4}{c}{$\wh{\bm\beta}^{\las}$} &		\multicolumn{4}{c}{$\wh{\bm\beta}^{\ds}$}
& \multicolumn{4}{c}{$\wh{\bm\beta}^{\las}$} &		\multicolumn{4}{c}{$\wh{\bm\beta}^{\ds}$}\\ \cmidrule(lr){4-7}\cmidrule(lr){8-11} \cmidrule(lr){12-15} \cmidrule(lr){16-19}
$d$ &	$n$ &	$p$ &	0 &	1 &	2 &	3 &	0 &	1 &	2 &	3  &	0 &	1 &	2 &	3 &	0 &	1 &	2 &	3   \\	\cmidrule(lr){1-3} \cmidrule(lr){4-7}\cmidrule(lr){8-11} \cmidrule(lr){12-15} \cmidrule(lr){16-19}
1 & 200 & 10 &  81  & 10   & 4   & 5  & 91  &  6  &  2 &   1 &  64  &  17   & 11    & 8  &  64   & 12    &16   &  8\\
 & 200 & 20 &  94  &  6  &  0 &   0  & 94  &  5 &   1  &  0 &68  &  10  &   9  &  13 &   75  &  10  &   7   &  8\\
 & 500 & 10 & 94  &  5 &   1&    0 &  86   & 7   & 4  &  3 &65 &   17  &  11  &   7   & 65  &  18 &    9 &    8\\ 
 & 500 & 20 &  97  &  2 &   0   & 1 &  98  &  1 &   1   & 0 &70 &   15  &   8   &  7  &  64  &  14  &  10  &  12\\
 \cmidrule(lr){1-19}
 &  &  & -2 & -1 & 0 & 1 & -2 & -1 & 0 & 1 & -2 & -1 & 0 & 1 & -2 & -1 & 0 & 1  \\   \cmidrule(lr){4-7}\cmidrule(lr){8-11} \cmidrule(lr){12-15} \cmidrule(lr){16-19}
3 & 200 & 10 &   0 &   0  & 77 &  23   & 0  &  0  & 78  & 22 & 27 &    3 &   49  &  21  &  30  &   6   & 49   & 15\\
 & 200 & 20 &  0  &  0  & 97 &   3  &  0  &  0  & 85  & 15 & 32   &  1  &  48  &  19  &  31 &    2 &   58   &  9\\
 & 500 & 10 & 0  &  0  & 76  & 24  &  0  &  0  & 83  & 17 &30  &   4  &  43   & 23   & 29 &    2  &  40  &  29\\ 
 & 500 & 20 &  0  &  0 &  74  & 26 &   0  &  0  & 97  &  3 &29  &   3   & 45   & 23  &  25  &   4 &   53  &  18\\
  \bottomrule
\end{tabular}}
\end{table}

\subsection{CLIME vs. ACLIME estimators}
\label{sec:sim:adaptive}

We compare the performance of the adaptive and non-adaptive estimators for the VAR innovation precision matrix $\bm\Delta$ and its impact on the estimation of $\bm\Omega$, the inverse of the long-run covariance matrix of the data (see \hyperref[sec:step:three]{Step 3}).
We generate $\bm\chi_t$ as in~\ref{m:ar}, fix $d = 1$ and treat it as known and consider $(n, p) \in \{(200, 50), (200, 100), (500, 100), (500, 200)\}$.

\begin{figure}[htbp]
\centering
\includegraphics[width = .8\textwidth]{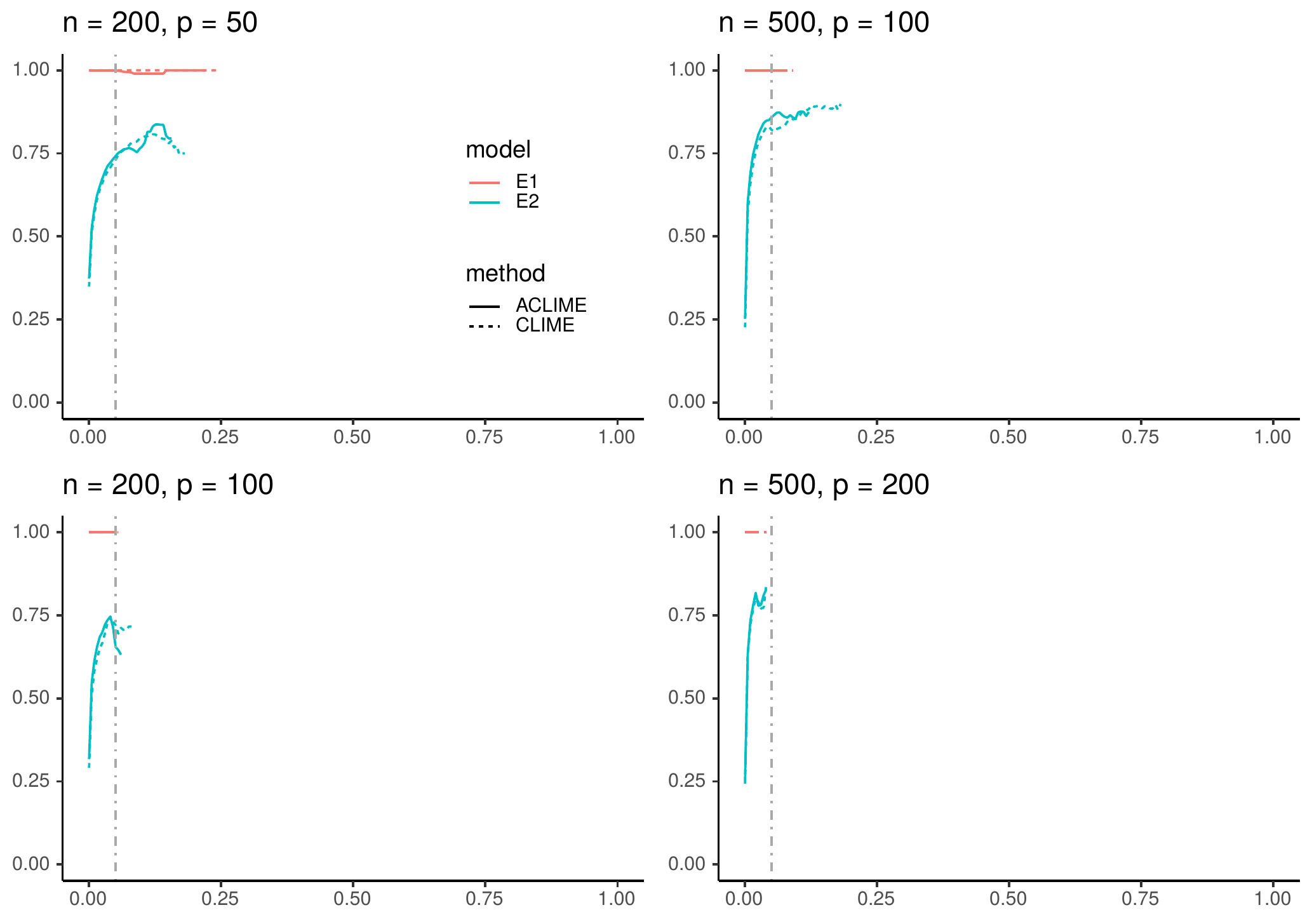}
\caption{\small ROC curves of TPR against FPR for
$\wh{\bm\Delta}$ with CLIME and ACLIME estimators
in recovering the support of $\bm\Delta$, averaged over $100$ realisations.
Vertical lines indicate FPR $= 0.05$.}
\label{fig:roc:delta}
\end{figure}

\begin{figure}[htbp]
\centering
\includegraphics[width = .8\textwidth]{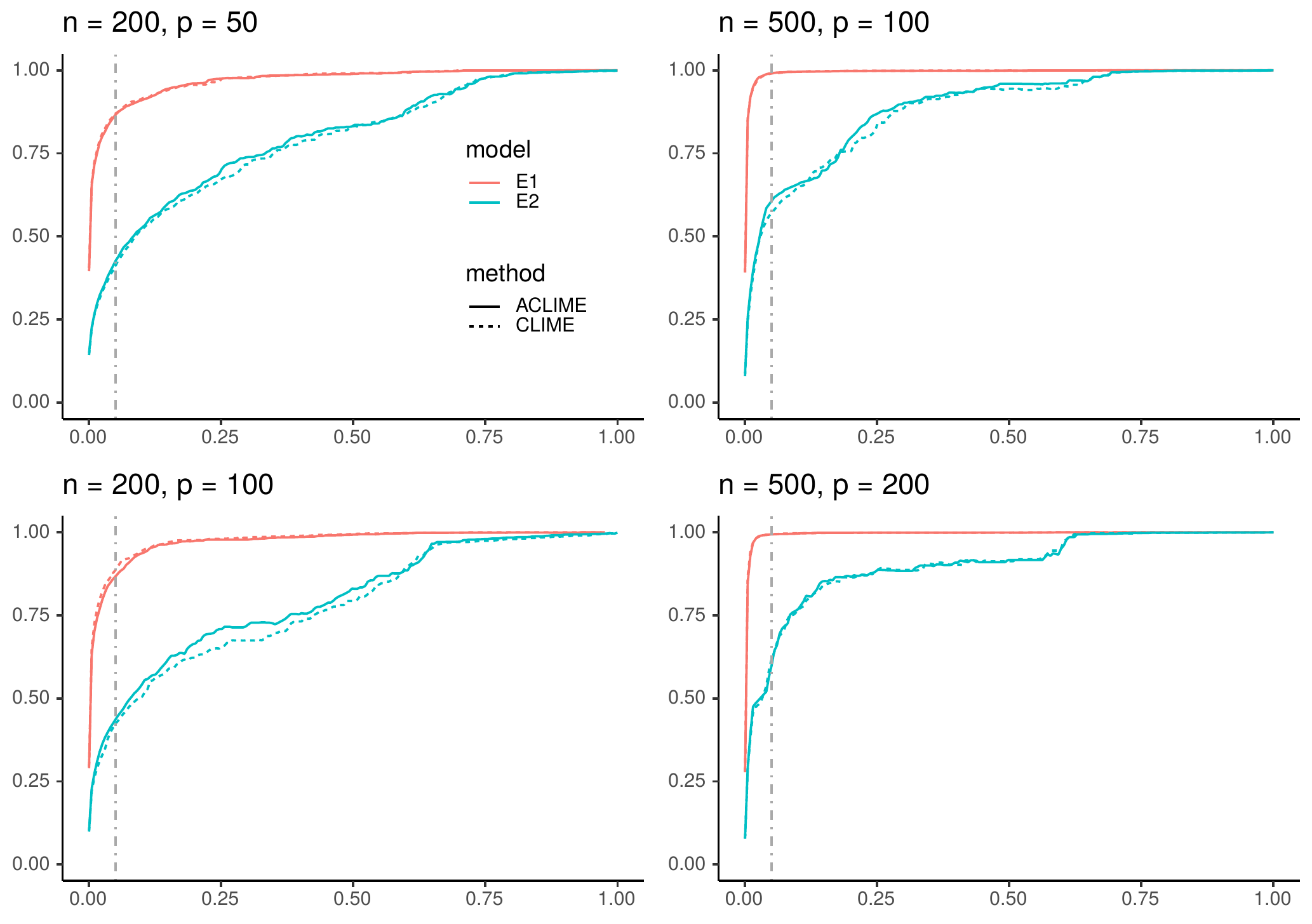}
\caption{\small ROC curves of TPR against FPR for $\wh{\bm\Omega}$ with CLIME and ACLIME estimators
in recovering the support of $\bm\Omega$, averaged over $100$ realisations. Vertical lines indicate FPR $= 0.05$.}
\label{fig:roc:omega}
\end{figure}

In Tables~\ref{table:lrpc:delta}~and~\ref{table:lrpc:omega}, we report the errors of $\bm\Delta$ and $\bm\Omega$.
We consider both the Lasso~\eqref{eq:lasso} and DS~\eqref{eq:ds} estimators of VAR parameters, and CLIME and ACLIME estimators for $\bm\Delta$, which lead to four different estimators for $\bm\Delta$ and $\bm\Omega$, respectively.
Overall, we observe that with increasing $n$, the performance of all estimators improve according to all metrics regardless of the scenarios~\ref{e:one} or~\ref{e:four}, while increasing $p$ has an adverse effect.
The two methods perform similarly in setting~\ref{e:one} when $\bm\Delta = \mbf I$.
There is marginal improvement for adopting the ACLIME estimator noticeable under~\ref{e:four}, particularly in TPR.
Figures~\ref{fig:roc:delta}~and~\ref{fig:roc:omega} shows the ROC curves for the support recovery of $\bm\Delta$ and $\bm\Omega$ when the Lasso estimator is used. 

\begin{table}[htbp]
\caption{Errors in estimating $\bm\Delta$ using CLIME and ACLIME estimators, measured by $L_F$ and $L_2$, averaged over $100$ realisations (with standard errors reported in brackets).
We also report the average TPR when FPR $= 0.05$ and the corresponding standard errors.} 
\label{table:lrpc:delta}
\centering
\resizebox{\columnwidth}{!}
{\scriptsize 
\begin{tabular}{ccc cccccc cccccc}
\toprule 
 &  &  & \multicolumn{6}{c}{CLIME} &  \multicolumn{6}{c}{ACLIME} \\
 \cmidrule(lr){4-9}\cmidrule(lr){10-15}
 &  &  & \multicolumn{3}{c}{$\wh{\bm\beta}^{\las}$} & \multicolumn{3}{c}{$\wh{\bm\beta}^{\ds}$} & \multicolumn{3}{c}{$\wh{\bm\beta}^{\las}$} & \multicolumn{3}{c}{$\wh{\bm\beta}^{\ds}$}   \\
 \cmidrule(lr){4-6}\cmidrule(lr){7-9} \cmidrule(lr){10-12} \cmidrule(lr){13-15}
Model & n & p & TPR & $L_F$ & $L_2$ & TPR & $L_F$ & $L_2$ & TPR & $L_F$ & $L_2$ & TPR & $L_F$ & $L_2$ \\
\cmidrule(lr){1-3}\cmidrule(lr){4-9} \cmidrule(lr){10-15}
\ref{e:one} & 200 & 50 & 1.000 & 0.215 & 0.489 & 1.000 & 0.220 & 0.497 & 1.000 & 0.207 & 0.472 & 1.000 & 0.209 & 0.469 \\
 &  &  & (0.000) & (0.047) & (0.223) & (0.000) & (0.047) & (0.182) & (0.002) & (0.043) & (0.173) & (0.000) & (0.041) & (0.116) \\
 & 200 & 100 & 1.000 & 0.235 & 0.513 & 1.000 & 0.241 & 0.521 & 1.000 & 0.223 & 0.507 & 1.000 & 0.228 & 0.518 \\
 &  &  & (0.000) & (0.036) & (0.089) & (0.000) & (0.036) & (0.107) & (0.000) & (0.033) & (0.084) & (0.000) & (0.034) & (0.099) \\
 & 500 & 100 & 1.000 & 0.181 & 0.458 & 1.000 & 0.183 & 0.466 & 1.000 & 0.176 & 0.452 & 1.000 & 0.178 & 0.458 \\
 &  &  & (0.000) & (0.022) & (0.062) & (0.000) & (0.029) & (0.087) & (0.000) & (0.022) & (0.052) & (0.000) & (0.028) & (0.069) \\
 & 500 & 200 & 1.000 & 0.198 & 0.510 & 1.000 & 0.193 & 0.492 & 1.000 & 0.187 & 0.505 & 1.000 & 0.182 & 0.489 \\
 &  &  & (0.000) & (0.027) & (0.066) & (0.000) & (0.035) & (0.065) & (0.000) & (0.026) & (0.056) & (0.000) & (0.033) & (0.057) \\	\cmidrule(lr){1-15}
 \ref{e:four} & 200 & 50 & 0.659 & 0.422 & 0.816 & 0.662 & 0.391 & 0.608 & 0.682 & 0.397 & 0.706 & 0.687 & 0.380 & 0.600 \\
 &  &  & (0.058) & (0.101) & (0.654) & (0.057) & (0.031) & (0.144) & (0.055) & (0.056) & (0.351) & (0.054) & (0.030) & (0.176) \\
 & 200 & 100 & 0.639 & 0.417 & 0.695 & 0.637 & 0.420 & 0.720 & 0.669 & 0.404 & 0.663 & 0.668 & 0.405 & 0.684 \\
 &  &  & (0.044) & (0.039) & (0.205) & (0.042) & (0.043) & (0.249) & (0.041) & (0.037) & (0.162) & (0.039) & (0.037) & (0.193) \\
 & 500 & 100 & 0.730 & 0.372 & 0.764 & 0.726 & 0.499 & 1.708 & 0.735 & 0.358 & 0.650 & 0.734 & 0.361 & 0.718 \\
 &  &  & (0.035) & (0.097) & (0.828) & (0.039) & (1.101) & (7.586) & (0.032) & (0.038) & (0.322) & (0.031) & (0.056) & (0.517) \\
 & 500 & 200 & 0.729 & 0.370 & 0.711 & 0.728 & 0.362 & 0.736 & 0.737 & 0.363 & 0.647 & 0.737 & 0.354 & 0.673 \\
 &  &  & (0.028) & (0.035) & (0.355) & (0.028) & (0.035) & (0.384) & (0.023) & (0.026) & (0.239) & (0.024) & (0.028) & (0.279)\\	\bottomrule
\end{tabular}}
\end{table}

\begin{table}[htbp]
\caption{Errors in estimating $\bm\Omega$ using CLIME and ACLIME estimators of $\bm\Delta$, measured by $L_F$ and $L_2$, averaged over $100$ realisations (with standard errors reported in brackets).
We also report the average TPR when FPR $= 0.05$ and the corresponding standard errors.} 
\label{table:lrpc:omega}
\centering
\resizebox{\columnwidth}{!}
{\scriptsize 
\begin{tabular}{ccc cccccc cccccc}
\toprule 
 &  &  & \multicolumn{6}{c}{CLIME} &  \multicolumn{6}{c}{ACLIME} \\
 \cmidrule(lr){4-9}\cmidrule(lr){10-15}
 &  &  & \multicolumn{3}{c}{$\wh{\bm\beta}^{\las}$} & \multicolumn{3}{c}{$\wh{\bm\beta}^{\ds}$} & \multicolumn{3}{c}{$\wh{\bm\beta}^{\las}$} & \multicolumn{3}{c}{$\wh{\bm\beta}^{\ds}$}   \\
 \cmidrule(lr){4-6}\cmidrule(lr){7-9} \cmidrule(lr){10-12} \cmidrule(lr){13-15}
Model & n & p & TPR & $L_F$ & $L_2$ & TPR & $L_F$ & $L_2$ & TPR & $L_F$ & $L_2$ & TPR & $L_F$ & $L_2$ \\
\cmidrule(lr){1-3}\cmidrule(lr){4-9} \cmidrule(lr){10-15}
\ref{e:one} & 200 & 50 & 0.871 & 0.415 & 0.557 & 0.862 & 0.422 & 0.571 & 0.867 & 0.411 & 0.558 & 0.856 & 0.417 & 0.570 \\
 &  &  & (0.108) & (0.050) & (0.070) & (0.119) & (0.055) & (0.080) & (0.106) & (0.051) & (0.088) & (0.114) & (0.053) & (0.083) \\
 & 200 & 100 & 0.883 & 0.432 & 0.589 & 0.896 & 0.438 & 0.595 & 0.868 & 0.423 & 0.583 & 0.883 & 0.429 & 0.587 \\
 &  &  & (0.084) & (0.050) & (0.072) & (0.080) & (0.046) & (0.065) & (0.088) & (0.048) & (0.077) & (0.085) & (0.045) & (0.061) \\
 & 500 & 100 & 0.991 & 0.331 & 0.492 & 0.989 & 0.339 & 0.499 & 0.991 & 0.328 & 0.490 & 0.989 & 0.337 & 0.498 \\
 &  &  & (0.016) & (0.031) & (0.069) & (0.021) & (0.036) & (0.065) & (0.015) & (0.033) & (0.070) & (0.019) & (0.036) & (0.067) \\
 & 500 & 200 & 0.994 & 0.352 & 0.529 & 0.992 & 0.351 & 0.540 & 0.994 & 0.344 & 0.525 & 0.990 & 0.342 & 0.537 \\
 &  &  & (0.009) & (0.038) & (0.054) & (0.018) & (0.045) & (0.065) & (0.009) & (0.038) & (0.056) & (0.014) & (0.044) & (0.068) \\	\cmidrule(lr){1-15}
 \ref{e:four} & 200 & 50 & 0.509 & 0.532 & 0.724 & 0.510 & 0.514 & 0.664 & 0.504 & 0.518 & 0.679 & 0.507 & 0.506 & 0.658 \\
 &  &  & (0.078) & (0.071) & (0.243) & (0.068) & (0.043) & (0.137) & (0.071) & (0.055) & (0.162) & (0.063) & (0.043) & (0.141) \\
 & 200 & 100 & 0.511 & 0.541 & 0.683 & 0.513 & 0.542 & 0.695 & 0.509 & 0.531 & 0.674 & 0.504 & 0.531 & 0.679 \\
 &  &  & (0.059) & (0.047) & (0.082) & (0.065) & (0.051) & (0.093) & (0.062) & (0.045) & (0.084) & (0.061) & (0.046) & (0.084) \\
 & 500 & 100 & 0.640 & 0.450 & 0.655 & 0.624 & 0.544 & 1.099 & 0.642 & 0.441 & 0.597 & 0.637 & 0.440 & 0.617 \\
 &  &  & (0.066) & (0.072) & (0.402) & (0.079) & (0.866) & (3.714) & (0.059) & (0.036) & (0.118) & (0.060) & (0.047) & (0.204) \\
 & 500 & 200 & 0.670 & 0.461 & 0.630 & 0.658 & 0.450 & 0.630 & 0.677 & 0.456 & 0.612 & 0.661 & 0.445 & 0.605 \\
 &  &  & (0.045) & (0.041) & (0.116) & (0.043) & (0.040) & (0.117) & (0.041) & (0.036) & (0.075) & (0.037) & (0.037) & (0.082)\\	\bottomrule
\end{tabular}}
\end{table}

\clearpage

\section{Appendix D: Dataset information}
\label{sec:real:data}

Table~\ref{table:definitions} defines the four node types in the panel.
Table~\ref{table:data:info} describes the dataset analysed in \hyperref[sec:real]{Data example}. 

\begin{table}[htb]
    \caption{Node type definitions for energy price data.}
    \label{table:definitions}
{\footnotesize
    \centering
    \begin{tabular}{c l}
    \toprule
    Name & Definition \\
    \cmidrule(lr){1-1} \cmidrule(lr){2-2} 
    Zone & A transmission owner's area within the PJM Region.  \\
    Aggregate & A group of more than one individual bus into a pricing node (pnode) \\\
    & that is considered as a whole in the Energy Market and other various systems \\
    & and Markets within PJM. \\
    Hub & A group of more than one individual bus into a regional pricing node (pnode) \\
    &developed to produce a stable price signal in the Energy Market \\
    &and other various systems and Markets within PJM.\\
    Extra High Voltage (EHV) & Nodes at 345kV and above on the PJM system. \\
    \bottomrule
    \end{tabular}}
\end{table}

\begin{table}[htb]
\caption{\small Names, IDs and Types for the $50$ power nodes in the energy price dataset.}
\label{table:data:info}
\centering
{ \small
\begin{tabular}{c c c}
\toprule 
Name &	Node ID 	&	Node Type	\\	
\cmidrule(lr){1-1} \cmidrule(lr){2-2} \cmidrule(lr){3-3}  		
PJM & 1 & ZONE \\
AECO & 51291 & ZONE \\
BGE & 51292 & ZONE \\
DPL & 51293 & ZONE \\
JCPL & 51295 & ZONE \\
METED & 51296 & ZONE \\
PECO & 51297 & ZONE \\
PEPCO & 51298 & ZONE \\
PPL & 51299 & ZONE \\
PENELEC & 51300 & ZONE \\
PSEG & 51301 & ZONE \\
\cmidrule(lr){1-1} \cmidrule(lr){2-2} \cmidrule(lr){3-3}  
BRANDONSH & 51205 & AGGREGATE \\
BRUNSWICK & 51206 & AGGREGATE \\
COOKSTOWN & 51211 & AGGREGATE \\
DOVER & 51214 & AGGREGATE \\
DPL NORTH & 51215 & AGGREGATE \\
DPL SOUTH & 51216 & AGGREGATE \\
EASTON & 51218 & AGGREGATE \\
ECRRF & 51219 & AGGREGATE \\
EPHRATA & 51220 & AGGREGATE \\
FAIRLAWN & 51221 & AGGREGATE \\
HOMERCIT & 51229 & AGGREGATE \\
HOMERCIT UNIT1 & 51230 & AGGREGATE \\
HOMERCIT UNIT2 & 51231 & AGGREGATE \\
HOMERCIT UNIT3 & 51232 & AGGREGATE \\
KITTATNY 230 & 51238 & AGGREGATE \\
MANITOU & 51239 & AGGREGATE \\
MONTVILLE & 51241 & AGGREGATE \\
PENNTECH & 51246 & AGGREGATE \\
PPL\_ALLUGI & 51252 & AGGREGATE \\
SENECA & 51255 & AGGREGATE \\
SOUTHRIV 230 & 51261 & AGGREGATE \\
SUNBURY LBRG & 51270 & AGGREGATE \\
TRAYNOR & 51277 & AGGREGATE \\
UGI & 51279 & AGGREGATE \\
VINELAND & 51280 & AGGREGATE \\
WELLSBORO & 51285 & AGGREGATE \\
\cmidrule(lr){1-1} \cmidrule(lr){2-2} \cmidrule(lr){3-3}  
EASTERN HUB & 51217 & HUB \\
WEST INT HUB & 51287 & HUB \\
WESTERN HUB & 51288 & HUB \\
\cmidrule(lr){1-1} \cmidrule(lr){2-2} \cmidrule(lr){3-3}  	
ALBURTIS & 52443 & EHV \\
BRANCHBURG & 52444 & EHV \\
BRIGHTON & 52445 & EHV \\
BURCHESHILL & 52446 & EHV \\
CALVERTC & 52447 & EHV \\
CHALKPT & 52448 & EHV \\
CONASTONE & 52449 & EHV \\
CONEMAUGH & 52450 & EHV \\
DEANS & 52451 & EHV \\
ELROY & 52452 & EHV
	\\
\bottomrule
\end{tabular}}
\end{table}
\end{article}

\end{document}